\documentclass[aps,twocolumn,showpacs]{revtex4}
\usepackage{dcolumn}
\usepackage{float}
\usepackage{graphicx}
\usepackage{amsmath}
\usepackage{verbatim}
\usepackage{amssymb}
\usepackage{psfrag}
\usepackage{color}
\usepackage{bm}
\usepackage{wrapfig}
\usepackage{subfigure}
\usepackage{makeidx}
\usepackage{bm}
\usepackage{epsf}
\usepackage{multirow}
\usepackage{mathrsfs}
\usepackage{braket}
\usepackage{xcolor}
\usepackage[colorlinks,linkcolor=red,urlcolor=blue,citecolor=blue]{hyperref}

\begin{document}
	\title{Dirac monopole magnets in non-Hermitian systems}
	\author{Haiyang Yu$^{1}$}
    \author{Tao Jiang$^{1}$}
	\author{Li-Chen Zhao$^{1,2,3}$}\email{zhaolichen3@nwu.edu.cn}
	\affiliation{$^{1}$School of Physics, Northwest University, Xi'an 710127, China}
	\affiliation{$^{2}$Shaanxi Key Laboratory for Theoretical Physics Frontiers, Xi'an 710127, China}
	\affiliation{$^{3}$NSFC-SPTP Peng Huanwu Center for Fundamental Theory, Xi'an 710127, China}
\date{\today}
	\begin{abstract}
We theoretically establish that non-Hermitian perturbations induce a topological transformation of point-like Dirac monopoles into extended monopole distributions, characterized by distinct charge configurations emergent from three distinct Berry connection forms. Using piecewise adiabatic evolution, we confirm the validity of these configurations through observations of complex geometric phases. Most critically, we find a quantitative relation $\Delta \phi_d = \Delta \phi_g$, which quantifies how  cumulative minute energy differences (\(\Delta \phi_d\)) manifest as  geometric phase shifts (\(\Delta \phi_g\)) uniquely in non-Hermitian systems. We further propose a scheme leveraging soliton dynamics in dissipative two-component Bose-Einstein condensates, enabling direct measurement of these topological signatures. These results establish a milestone for understanding Dirac monopole charge distributions and measuring complex geometric phases in non-Hermitian systems, with far-reaching implications for topological quantum computing and non-Hermitian photonics.
	\end{abstract}
\maketitle

\section{Introduction}
P. A. M. Dirac first predicted the existence of magnetic monopoles in real space, drawing on the zeros of wave functions and their properties related to non-integrable phase factors \cite{Dirac}. Through studies of geometric phase variations, M. V. Berry demonstrated that a virtual Dirac monopole vector potential field surrounds energy degeneracy points in parameter space \cite{Berry1984}. Since then, numerous point-like virtual monopoles have been identified in the parameter spaces of Hermitian systems \cite{gphase1,RMP1,Niu,RMP2}, while nonlinear systems can host magnetic monopoles with alternative morphologies, such as line-shaped or disk-shaped configurations \cite{WuLiu,Liu}. Monopole charges and their distributions fundamentally govern topological effects \cite{topology1,topology2,topology3,Ding,Herm,Leykam,DSBorgnia,ZSYang,HGZirnstein}, geometric phases \cite{gphase1,RMP1,Niu,RMP2,WuLiu,Liu}, as well as a variety of anomalous transport phenomena and dynamical behaviors in both matter and optical systems \cite{AChen,Yin,TGRappoport,DDreon,GQXu1,NonHDirac}.

\begin{figure}[t]
	\centering
	\includegraphics[width=8.7cm,height=6.3cm]{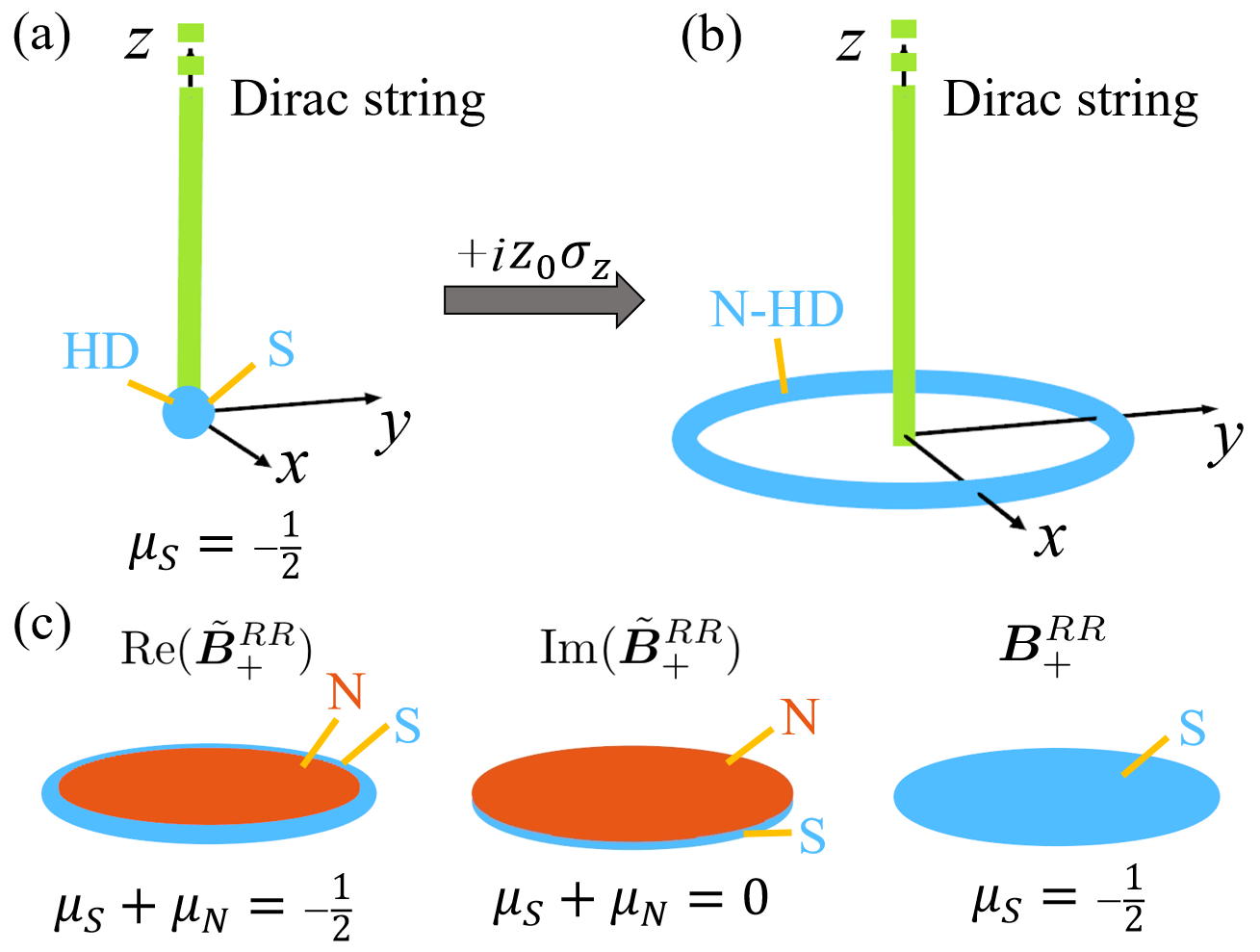}
	\caption{\label{1}(a) The magnetic monopole and Dirac string of $\ket{\psi^R_+(\bm{R})}$ when the system is Hermitian ($Z_0=0$). (b) The degenerate ring and Dirac string of $\ket{\psi^R_+(\bm{R})}$ when the system is Hermitian ($Z_0=1$). (c) The monopole disks of $\bm{\tilde{B}}^{RR}_+$ and $\bm{B}^{RR}_+$ for the non-Hermitian system. HD denotes the degenerate point of energy spectrum for a Hermitian system, and N-HD denotes the degenerate point of energy spectrum for a non-Hermitian system. Blue color denotes S charges, red color denotes N charges. $\mu_S$ and $\mu_N$ denote S and N charges, respectively. }
\end{figure}

Dirac's original monopole theory posits that monopoles are invariably situated at the endpoints of nodal lines \cite{Dirac}. Moreover, within the framework of the Berry phase, it is generally anticipated that monopoles should lie at points of energy degeneracy \cite{Berry1984}. Notably, our analysis reveals that in non-Hermitian systems, energy degeneracies do not necessarily coincide with the endpoints of Dirac strings (see Fig. \ref{1})---a correspondence that holds robustly in Hermitian systems \cite{Zhao}.
While monopole charges and their distributions can be effectively characterized using topological vector potentials \cite{Dirac}, non-Hermitian systems exhibit several distinct forms of such vector potentials, commonly referred to as Berry connections \cite{Xu,Shen,Silberstein,Wu}. To the best of our knowledge, the fundamental relationships and differences between these forms have not yet been fully elucidated, which makes it be hard to use them properly for topology studies in non-Hermitian systems. These distinctive features underscore the need for a more fundamental investigation into Berry connections in non-Hermitian systems.

In this paper, we derive and investigate three distinct forms of Berry connections ($\bm{A}^{LR}$, $\bm{\tilde{A}}^{RR}$, and $\bm{A}^{RR}$) in non-Hermitian systems. Our approach combines foundational principles of adiabatic processes with direct numerical computations. We show that the Chern number for linear bands of non-Hermitian systems can also be obtained without bi-orthogonality, for which the Berry connection $\bm{\tilde{A}}^{RR}$ expressed by pure right eigenstates.  We find that non-Hermitian terms can transform a Dirac monopole point into a monopole magnet, and different Berry connection forms exhibit distinct monopole charge distributions. Although the total charge is identical for all forms, these distinct distributions lead to different geometric phase variations. The validity of these forms is confirmed by observing complex geometric phases through piecewise adiabatic evolution.
We demonstrate that only the $\bm{A}^{RR}$ form is valid when the dynamical phase is calculated using time-dependent expectation value of energy. In contrast, the other two forms ($\bm{A}^{LR} = \bm{\tilde{A}}^{RR}$) are valid when the dynamical phase is defined via the eigenvalues of the instantaneous eigenequation. While the discrepancy between the energy expectation value and the instantaneous eigenvalue is negligible in Hermitian systems, it induces significant geometric phase effects in non-Hermitian systems.
Notably, we uncover a general relationship between the accumulation of small energy differences ($\Delta \phi_d$) and the difference in geometric phases ($\Delta \phi_g$) derived from the distinct Berry connection forms ($\bm{\tilde{A}}^{RR}$ and $\bm{A}^{RR}$). Specifically, the relation $\Delta \phi_d = \Delta \phi_g$ holds for both Hermitian and non-Hermitian systems, with $\Delta \phi_d = \Delta \phi_g=0$ always standing for Hermitian systems.  Our results significantly deepen the understanding of Dirac monopole charge distributions and offer key insights for measuring complex geometric phases in non-Hermitian systems.

\section{Different Berry connections of non-Hermitian systems}
Recently, numerous novel physical phenomena \cite{rmp1,Ueda,NOkuma,np1,Herm2,prx1,np3,sc1,nature1,prl1} have been widely observed in non-Hermitian systems, closely related to the non-orthogonality of states and their coincidence at degeneracy points \cite{book1,book2,book3}. However, several fundamental aspects of topology involving Dirac monopoles and their related topological effects remain debated, due to incomplete understanding of the distinct forms of vector potentials. This necessitates exploring different vector potential (Berry connection) forms in non-Hermitian systems from more fundamental perspectives.
For simplicity without losing generality, we consider a general $2\times 2$ non-Hermitian models $H$ with a parameters space $\bm{R}$ to discuss the Dirac monopoles and their associated geometric phases, where $\bm{R}$ is the control parameter that governs the system's dependence on $t$.

We derive Berry connections by analyzing the evolution of initial eigenstates under adiabatic processes. The wave function can be expressed as $\ket{\psi}=C_1(t)\ket{\psi^R_+(\bm{R})}e^{i\gamma_+ (\bm{R})}e^{i \phi_+(t)}+C_2(t)\ket{\psi^R_-(\bm{R})}e^{i\gamma_- (\bm{R})}e^{i\phi_-(t)}$, where $C_{1, 2}(t)$, $\ket{\psi_\pm^R(\bm{R})}$ and $\phi_\pm(t)$ respectively denote expansion coefficients, instantaneous eigenstates, and dynamical phases. The instantaneous states are given by the instantaneous eigen-equation $H(\bm{R}) \ket{\psi^R_{\pm}(\bm{R})} =\ E_{\pm}(\bm{R})  \ket{\psi^R_{\pm}(\bm{R})}$. It has a conjugated partner, which is described by $H^{\dag}(\bm{R}) \ket{\psi^L_{\pm}(\bm{R})} =\ E^{*}_{\pm}(\bm{R})  \ket{\psi^L_{\pm}(\bm{R})}$. Additionally, the non-integrable phase factor $\gamma_\pm(\bm{R})$ should be introduced due to the presence of Dirac string with endpoint or energy degeneracy \cite{Dirac,Berry1984}.
By substituting the expansion form into the time-dependent Schr\"odinger equation, one can has different orders of adiabatic parameter. The zero order of $\frac{d \bm{R}} {dt}$ gives dynamical phases, while the first order describes the geometric phases (non-integrable phase factor).  By multiplying different eigenstates, distinct
forms of Berry connections are obtained (see details in Appendix A). The conventional form is:
\begin{eqnarray}\label{ALR}
\bm{A}^{LR}_\pm&=&i\frac{\bra{\psi^L_\pm}\bigtriangledown_{\bm{R}}\ket{\psi^R_\pm}}
{\braket{\psi^L_\pm|\psi^R_\pm}},
\end{eqnarray}
where $\bra{\psi^L_\pm(\bm{R})}$ and $\ket{\psi^R_\pm(\bm{R})}$ are abbreviated as $\bra{\psi^L_\pm}$ and $\ket{\psi^R_\pm}$.
Eq. (\ref{ALR}) involves both left and right eigenstates \cite{Garrison}. The form is indeed simple due to the bi-orthogonality of states. Quantized indices of complex bands are typically obtained using bi-orthogonal relations \cite{rmp1,Garrison,Ueda,NOkuma,np1,Herm2}. Here, we show that the Chern number for linear bands can also be obtained without bi-orthogonality. We derive the Berry connection using pure right eigenstates as:
\begin{equation}\label{ARR}
\bm{\tilde{A}}^{RR}_\pm=i\frac{\bra{\psi_\pm^R}\bigtriangledown_{\bm{R}} \ket{\psi_\pm^R}\braket{\psi_\mp^R|\psi_\mp^R}
	-\bra{\psi_\mp^R}\bigtriangledown_{\bm{R}}\ket{\psi_\pm^R}\braket{\psi_\pm^R|\psi_\mp^R}}
{\braket{\psi_\pm^R|\psi_\pm^R}\braket{\psi_\mp^R|\psi_\mp^R}
	-\braket{\psi_\mp^R|\psi_\pm^R}\braket{\psi_\pm^R|\psi_\mp^R}}.
\end{equation}
Under adiabatic conditions, these two forms of Berry connections are equivalent for generic two-level non-Hermitian systems, namely, $\bm{A}^{LR}_\pm=\bm{\tilde{A}}^{RR}_\pm $.

Another form of Berry connection using right vectors is \cite{Xu,Shen,Silberstein,Wu}:
\begin{eqnarray}
\bm{A}^{RR}_\pm&=&i\frac{\bra{\psi^R_\pm}\bigtriangledown_{\bm{R}}\ket{\psi^R_\pm}}
{\braket{\psi^R_\pm|\psi^R_\pm}},
\end{eqnarray}
which is analogous to the Berry connection in Hermitian systems \cite{Berry1984}. Notably, this form cannot be explicitly derived for non-Hermitian models. For Hermitian systems, $\bm{\tilde{A}}^{RR}_\pm=\bm{A}^{RR}_\pm$, but they generally differ in non-Hermitian systems due to the non-orthogonality of eigenstates. Other common forms $\bm{A}^{RL}_\pm$ and $\bm{\tilde{A}}^{LL}_\pm$ can be derived similarly by analyzing the evolution of left states (eigenstates of $H^{\dag}$).

The Berry curvature (effective magnetic field) in parameter space is obtained by taking the curl of the Berry connection. In non-Hermitian systems, the magnetic fields $\bm{\tilde{B}}^{RR}_\pm$ and $\bm{B}^{RR}_\pm$ are generally distinct, despite their identical total magnetic charges \cite{Shen,Silberstein,Nakamura,Jezequel}. This implies different geometric phases derived from the two connections:
\begin{eqnarray}
	\oint_C \bm{\tilde{A}}^{RR}_\pm \cdot d \bm{R}= \oint_C \bm{A}^{LR}_\pm \cdot d \bm{R} , \label{vt}
\end{eqnarray}
\begin{eqnarray}
	\oint_C \bm{A}^{RR}_\pm \cdot d \bm{R}. \label{motion}
\end{eqnarray}	
The validity and measurability of these forms in experiments can be verified by direct numerical simulations of geometric phases.

\section{An explicit two-level system}
To explicitly illustrate the differences and test the validity of these forms, we consider a two-level system:
$H(\bm{R})=\begin{pmatrix}
	Z+iZ_0&X-iY\\
	X+iY&-Z-iZ_0
\end{pmatrix}$, where $\bm{R}=(X,Y,Z)$, $Z_0$ describes the non-Hermitian effects.
The eigenvalues of the Hamiltonian are $E_\pm(\bm{R})=\pm\sqrt{X^2+Y^2+Z^2-Z_0^2+2iZZ_0}$. 
The degeneracies of the entire energy spectrum lie at $Z=0$, $X^2+Y^2=Z_0^2$, which are typically referred to as non-Hermitian degeneracies \cite{Berry2} or exceptional points \cite{EPR,EPR2}.
For such a non-Hermitian system, its parameter space forms a Riemannian surface \cite{Wu,IIArkhipov,KWang,HNasari}, and different methods of cutting this Riemannian surface result in distinct selections of eigenstates \cite{Wu}.
From the perspective of physical measurements, it is considered more appropriate to mark the different eigenvalues by the real part of the energy spectrum \cite{QZhong}, as their differences are easily measurable via optical spectroscopic measurements.
Based on this, we denote $E$ as $E_\pm=\pm(a+ib)$, where $a=\sqrt{\frac{X^2+Y^2+Z^2-Z_0^2+\sqrt{(X^2+Y^2+Z^2-Z_0^2)^2+4(ZZ_0)^2}}{2}}$, $b=\frac{ZZ_0}{a}$. Accordingly, the eigenstates of the system can be expressed as $
\ket{\psi^R_\pm(\bm{R})}=
\begin{pmatrix}
	X-iY\\
	-Z-iZ_0+E_\pm
\end{pmatrix}$.
The Berry curvature corresponding to each eigenstate can be derived, as detailed in Appendix A.

According to Berry's framework \cite{Berry1984}, monopoles generally locate at energy degeneracies. On the other hand, the monopoles should lie at the endpoints of Dirac strings, which can known by analyzing the nodal lines of the eigenstates  \cite{Dirac,Zhao}.
For the Hermitian case ($Z_0=0$),  with the state $\ket{\psi^R_+(\bm{R})}$ a $-\frac{1}{2}$ magnetic charge exists at the energy degeneracy. The endpoint of Dirac string and energy degeneracy locate at identical positions in parameters space, shown in Fig. \ref{1} (a). However, this consistency may be disrupted in non-Hermitian systems. For non-Hermitian cases ($Z_0\neq 0$),  the endpoint of Dirac string and energy degeneracy obviously do not admit identical position anymore, as shown in Fig. \ref{1} (b).

\begin{figure}[t]
	\centering
	\includegraphics[width=8.7cm,height=3.5cm]{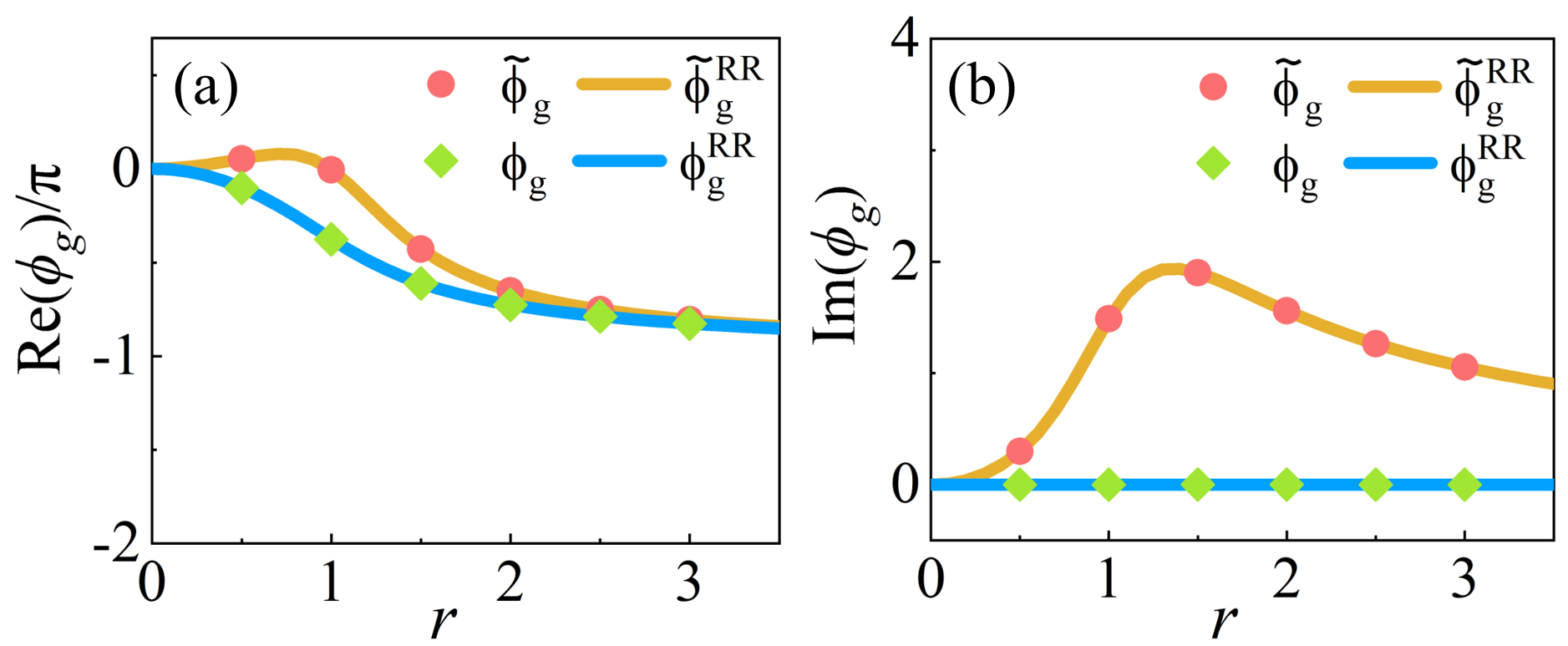}
	\caption{\label{2}(a) The theoretical and numerical comparison diagram of the real part of the Berry phase in the initial state $\ket{\psi^R_+(\bm{R})}$ after one period of evolution in different orbits. (b) The theoretical and numerical comparison diagram of the imaginary part of the Berry phase in the initial state $\ket{\psi^R_+(\bm{R})}$ after one period of evolution in different orbits. The specific information of the evolution orbit is: $Z=0.5$, $X=r\cos(\omega t)$, $Y=r\sin(\omega t)$, $r=\sqrt{X^2+Y^2}$, $\omega=0.0005\pi$ and the orbit is $\omega t: 0\to 2\pi$. The non-Hermitian parameter is $Z_0=1$.  The difference of orbitals is reflected in the difference of $r$. This figure statistics the change of Berry phase when $r$ goes from 0 to 3.5.}
\end{figure}
Notably, monopole charges are distributed not only at the endpoint of the Dirac string and the non-Hermitian degeneracy, but also in the entire degeneracy region of the real part of the energy spectrum. Fig. \ref{1} (c) shows the monopole charges given by $\bm{\tilde{B}}^{RR}_+$ and $\bm{B}^{RR}_+$ of $\ket{\psi^R_+(\bm{R})}$ ($Z_0=1$), where $\bm{\tilde{B}}^{RR}_+$ is a complex magnetic field and $\bm{B}^{RR}_+$ has only the real part. The disk of $\mathrm{Re}(\bm{\tilde{B}}^{RR}_+)$ contains N charges internally and S charges at the edge, with more S charges $\mu_S$ than N charges $\mu_N$. the total charge $\mu_S+\mu_N$ equals the S monopole charge in the Hermitian case ($\mu_S=-1/2$ for $Z_0=0$). We term this a ``semi-monopole magnet" to distinguish it from previous monopoles \cite{RMP1,Niu,RMP2,WuLiu} and ordinary magnets.
Unlike $\bm{\tilde{B}}^{RR}_+$, the $\bm{B}^{RR}_+$ diskcontains only S charges, as shown in Fig. \ref{1} (c). Additionally, the $\mathrm{Im}(\bm{\tilde{B}}^{RR}_+)$ disk has an N pole at the top and an S pole at the bottom, with equal but opposite charges. However, differences in the magnetic field will result in differences in the geometric phase. It is therefore necessary to explore which Berry connection aligns with experimental measurements.

\section{A striking relation $\Delta \phi_d=\Delta \phi_g$}
To confirm the practical validity of the Berry connections, we perform numerical experiments to investigate variations of Berry phase under adiabatic conditions. Notably, distinct eigenstates of non-Hermitian systems exhibit adiabaticity in different parameter space regions \cite{Muga,Nenciu,Dridi,Wu19}.
For the model $H(\bm{R})$, adiabaticity holds well in the upper half-space ($Z>0$) for the $\ket{\psi^R_+(\bm{R})}$, but breaks down for $\ket{\psi^R_-(\bm{R})}$. The lower half-space ($Z<0$) only supports adiabaticity for  $\ket{\psi^R_-(\bm{R})}$.  Using $\ket{\psi^R_+(\bm{R})}$ as the initial state, we fix $Z=0.5$ and $Z_0=1$ for adiabatic evolution, with $X=r\cos(\omega t)$, $Y=r\sin(\omega t)$ ($r=\sqrt{X^2+Y^2}$) and evolution trajectory is: $\omega t: 0\to 2\pi$. A small $\omega$ ensures adiabaticity, with state fidelity $f_+(t)=|\braket{\psi(t)|\psi_+^R(\bm{R})}|^2 \approx 1$ throughout evolution.
Generally, the geometric phase $\phi_g$ is the difference between the total phase $\phi_{total}$ and the dynamic phase $\phi_{d}$: $\phi_g=\phi_{total}-\phi_d$.
Interestingly, the dynamical phase has two distinct calculation methods:
\begin{eqnarray}
\phi_{d}&=&-\int_{0}^{T}E(t)dt, \\
\tilde{\phi}_{d}&=&  -\int_{0}^{T}E_+(\bm{R})dt,
\end{eqnarray}
where
$E(t)=\frac{\bra{\psi(t)}H\ket{\psi(t)}}{\braket{\psi(t)|\psi(t)}}$ is the time-dependent energy expectation value, and $E_+(\bm{R})$ is the instantaneous eigenvalue from $
  H(\bm{R}) \ket{\psi^R_+(\bm{R})} =\ E_+(\bm{R})  \ket{\psi^R_+(\bm{R})}$.

\begin{figure}[t]
	\includegraphics[width=8.5cm,height=7.9cm]{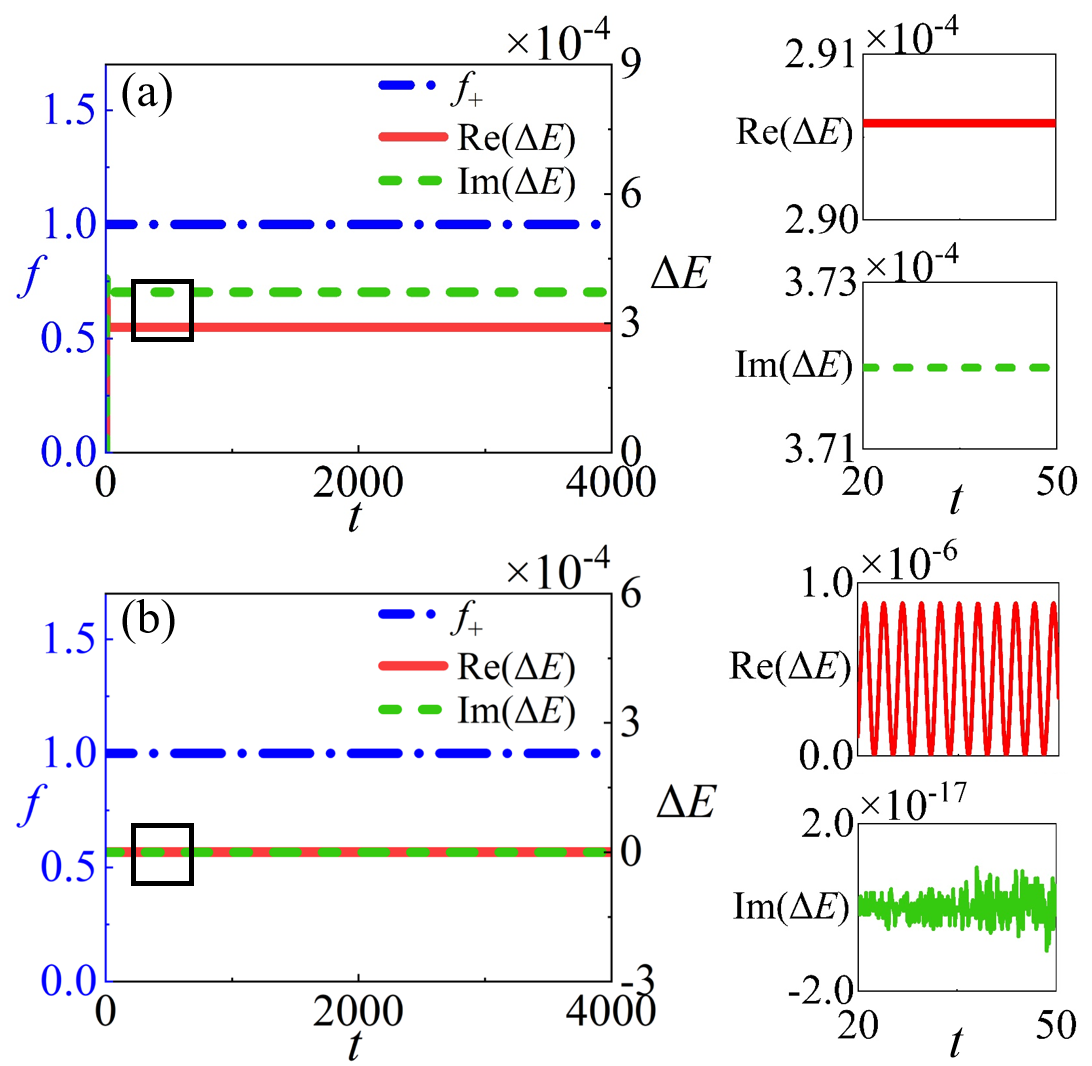}
	\caption{\label{3}(a) Evolution curve of fidelity and $\Delta E$ of non-Hermitian systems with time for $Z_0=1$, $Z=0.5$, $r=1$, $\omega=0.0005\pi$. (b) Evolution curve of fidelity and $\Delta E$ of Hermitian systems with time for $Z_0=0$, $Z=0.5$, $r=1$, $\omega=0.0005\pi$.}
\end{figure}

Most previous theoretical and experimental results use $\tilde{\phi}_d=-\int_{0}^{T}E_+(\bm{R}(t))dt$ predict complex geometric phases in non-Hermitian systems \cite{Garrison}, which have been experimentally measured  \cite{YSinghal}. This is reasonable for Hermitian systems but not generally valid for non-Hermitian systems. Our numerical experiments calculate the real and imaginary parts of $\phi_g$ and $\tilde{\phi}_g$ after one evolution cycle for $r$ ranging from $0$ to $3.5$, comparing them with the theoretical values $\phi_g^{RR}$ and $\tilde{\phi}_g^{RR}$ from the two Berry connections, as shown in Fig. \ref{2} (a) and (b). The results in the figure indicate that the geometric phases $\phi_g$ and $\tilde{\phi}_g$ correspond to those derived from $\phi_g^{RR} = \oint_C \bm{A}_+^{RR} \cdot d \bm{l}$ and $\tilde{\phi}_g^{RR} = \oint_C \bm{\tilde{A}}_+^{RR} \cdot d \bm{l}$. Namely,
\begin{eqnarray}\label{timedep4}
  \phi_g&=&\phi_{total}-\phi_{d}=\phi_g^{RR}
\end{eqnarray}
and
\begin{eqnarray}\label{timedep5}
  \tilde{\phi}_g &=&\phi_{total}-\tilde{\phi}_{d}=\tilde{\phi}_g^{RR}.
\end{eqnarray}
Explicitly, $\phi_g^{RR}$ and $\tilde{\phi}_g^{RR}$ exhibit distinct real and imaginary parts for the model. For example, $\phi_g^{RR}$ has a zero imaginary part, while $\tilde{\phi}_g^{RR}$ has a non-zero imaginary part (see Fig. \ref{2}  (b)), indicating a close relation between geometric phase differences and dynamical phase differences. Although the difference between $E(t)$ and $E_+(\bm{R})$ is extremely small, it can accumulate into an observable effect even in adiabatic processes for non-Hermitian systems.
Combining Eq. (\ref{timedep4}) and (\ref{timedep5}), we find
\begin{eqnarray}\label{deltad}
	\Delta \phi_g &=& \Delta \phi_d \nonumber\\
	&=&  \int_{0}^{T}\Delta E(t) dt,
\end{eqnarray}
where   $\Delta \phi_d=\phi_{d} -\tilde{\phi}_d$, $\Delta \phi_g= \tilde{\phi}_g^{RR}-\phi_g^{RR}$, and $\Delta E(t)=E_+(\bm{R})-E(t)$. This relation is tested directly with varying different parameters (see details in Appendix B).
This is one of the most fundamental results in this paper. In particular, this relation is derived analytically for an arbitrary $2\times2$ non-Hermitian model under the adiabatic approximation (see details in Appendix C). Notably, this relation also holds for Hermitian systems, where $\Delta \phi_g=\Delta \phi_d=0$. It directly indicates that the three different Berry connections become identical for Hermitian systems.

The energy difference warrants further investigation in light of Eq. (\ref{deltad}). Therefore, we will focus on the variation of $\Delta E(t)$ over the evolution process. Fidelity and energy difference were recorded over the evolution process, as shown in Fig. \ref{3} (a). Clearly, $f_+$ is extremely close to $1$, and $\Delta E(t)$ remains small throughout.
However, during the evolution process, $\Delta E(t)$ rapidly transitions from $0$ to a stable value and accumulates continuously, eventually reaching the same order of magnitude as the geometric phase. Thus, the small difference between $E(t)$ and $E_+(\bm{R})$ accumulates to observable effects in non-Hermitian systems. Importantly, Eq. (\ref{deltad}) holds under adiabatic conditions regardless of the evolution rate (see details in Appendix B) indicating that $\Delta \phi_d$ is fundamentally geometric rather than dynamical.
This differs significantly from Hermitian systems. Fig. \ref{3} (b) shows that the energy deviation during evolution is much smaller in Hermitian systems and oscillates, resulting in a negligible accumulated energy difference compared to the geometric phase.

 \section{Possibilities to observe the effects of monopole magnets}
 The effects of monopole magnets can be deduced from variations in the geometric phase by selecting distinct cyclic paths within the parameter space. Under weak nonlinear interactions, soliton states can be engineered to enhance the precision of geometric phase observations in experiments. This advantage arises from solitons' inherent stability---a result of the delicate balance between nonlinearity and dispersion---coupled with their flexibility in dynamic phase manipulation. Thus we propose a feasible experimental scheme to observe geometric phases and check the relation $\Delta \phi_d=\Delta \phi_g$  based on soliton evolution in two-component Bose-Einstein condensates (see details in  Appendix D).

\section{Conclusion and discussion}
We uncover that non-Hermitian effects can induce monopole magnets, by deriving distinct Berry connection forms ($\bm{A}^{LR}$, $\bm{\tilde{A}}^{RR}$, and $\bm{A}^{RR}$) based on fundamental adiabatic principles and direct numerical calculations. This challenges the conventional view of monopoles as localized at energy degeneracies, as shown in Fig. \ref{1} (c) where charges spread across parameter space rather than cluster at points. Through distinct Berry connections, we identify ``semi-monopole magnets" and disk-shaped monopole magnets. Notably, semi-monopole magnets host both inverse charges yet exhibit nonvanishing total charges, a feature absent in conventional monopoles \cite{RMP1,Niu,RMP2,WuLiu} and ordinary magnets.
The validity of these Berry connection forms is confirmed by observing complex geometric phases under piecewise adiabatic conditions. Our results clarify their scope of applicability: $\bm{A}^{RR}$ is reasonable when the dynamical phase is computed using the time-dependent expectation value of energy, while
 $\bm{A}^{LR}$ and $\bm{\tilde{A}}^{RR}$ are valid when the dynamical phase is determined by eigenvalues of instantaneous eigenequations. The relation $\Delta \phi_d = \Delta \phi_g$ establishes a quantitative link between energy discrepancies and geometric phases: specifically, the dynamical phase difference
 $\Delta \phi_d$ in adiabatic cyclic processes satisfies
\begin{eqnarray}
	\Delta \phi_d &=& \oint_C (\bm{\tilde{A}}^{RR} - \bm{A}^{RR}) \cdot d \bm{l} \nonumber\\
  &=& \oint_C (\bm{A}^{LR} - \bm{A}^{RR}) \cdot d \bm{l}.
\end{eqnarray}
This reflects another effect of the anomalous Berry connection effect in non-Hermitian systems \cite{Silberstein,WangXu}.  This work not only resolves some longstanding ambiguities in non-Hermitian Berry connection formalisms but also represents a key step toward establishing a foundational framework for quantifying topological charges and geometric phases in open quantum systems.

Our discussions show that non-Hermitian terms transform conventional Dirac monopole points into monopole magnets, similar to nonlinear systems \cite{WuLiu,Liu}, hinting at intrinsic relations between nonlinear Hermitian and linear non-Hermitian systems. Our results provide a unified understanding of Dirac monopole charge distributions and geometric phase measurements in non-Hermitian systems, promoting fundamental studies on monopoles and geometric phases in nonlinear non-Hermitian systems \cite{NN,Hu}. This advances non-Hermitian physics by unifying Dirac monopole theory with geometric phase measurements, with implications for quantum transport, topological phases, and dissipative quantum systems.

\section*{Acknowledgments}
This work is supported by the National Natural Science Foundation of China (Contracts No. 12375005, No. 12235007 and No. 12247103).

\begin{widetext}
\section*{Appendix A: Berry connections for two-level non-Hermitian systems}
We consider a general two-level non-Hermitian system, which is governed by parameters denoted by an abstract space $\bm{R}$, to derive the different Berry connections from basic principles. The Hamiltonian of the general system is denoted as $H(\bm{R})$, where $\bm{R}$ is a function of time $t$. The instantaneous eigenstates satisfy the instantaneous eigenvalue equation: $H(\bm{R}) \ket{\psi^R_{\pm}(\bm{R})} =\ E_{\pm}(\bm{R})  \ket{\psi^R_{\pm}(\bm{R})}$. These instantaneous eigenstates have conjugate counterparts, which are described by the equation: $H^{\dag}(\bm{R}) \ket{\psi^L_{\pm}(\bm{R})} =\ E^{*}_{\pm}(\bm{R})  \ket{\psi^L_{\pm}(\bm{R})}$.  An arbitrary state of the system can be expressed in the following form: $\ket{\psi(t)}=C_1(t)\ket{\psi^R_+(\bm{R})}e^{i\gamma_+ (\bm{R})}e^{i\phi_+(t)}+C_2(t)\ket{\psi^R_-(\bm{R})}e^{i\gamma_- (\bm{R})}e^{i\phi_-(t)}$, where $C_{1, 2}(t)$are the expansion coefficients, $\ket{\psi_\pm^R(\bm{R})}$ represent the instantaneous eigenstates and $\phi_\pm(t)$ are the dynamical phases, respectively. The non-integrable phase factors $\gamma_\pm(\bm{R})$ are introduced considering the existence of Dirac strings with endpoints or energy degeneracies\cite{Dirac,Berry1984}. These phase factors play an important role in characterizing the geometric properties of the system and cannot be ignored in the study of the system's quantum behavior.

By substituting the expansion form into the Schr\"odinger equation $i\frac{\partial \ket{\psi(t)}}{\partial t}=H(\bm{R})\ket{\psi(t)}$, the evolution equation of the coefficients can be obtained:
\begin{eqnarray}\label{Eq1}
		&&i\dot{C_1}(t)\ket{\psi_+^R(\bm{R})}e^{i\gamma_+(\bm{R})}e^{i\phi_+(t)}
		+i\dot{C_2}(t)\ket{\psi_-^R(\bm{R})}e^{i\gamma_-(\bm{R})}e^{i\phi_-(t)}\nonumber\\
		&&+iC_1(t)\frac{d\ket{\psi_+^R(\bm{R})}}{d\bm{R}}\frac{d\bm{R}}{dt}
		e^{i\gamma_+(\bm{R})}e^{i\phi_+(t)}
		+iC_2(t)\frac{d\ket{\psi_-^R(\bm{R})}}{d\bm{R}}\frac{d\bm{R}}{dt}
		e^{i\gamma_-(\bm{R})}e^{i\phi_-(t)}\nonumber\\
		&&-C_1(t)\frac{d\gamma_+(\bm{R})}{d\bm{R}}\frac{d\bm{R}}{dt}
		\ket{\psi^R_+(\bm{R})}e^{i\gamma_+(\bm{R})}e^{i\phi_+(t)}
		-C_2(t)\frac{d\gamma_-(\bm{R})}{d\bm{R}}\frac{d\bm{R}}{dt}
		\ket{\psi^R_-(\bm{R})}e^{i\gamma_-(\bm{R})}e^{i\phi_-(t)}\nonumber\\
		&&-C_1(t)\dot{\phi_+}(t)\ket{\psi^R_+(\bm{R})}
		e^{i\gamma_+(\bm{R})}e^{i\phi_+(t)}
		-C_2(t)\dot{\phi_-}(t)\ket{\psi^R_-(\bm{R})}
		e^{i\gamma_-(\bm{R})}e^{i\phi_-(t)}\nonumber\\
		&&=C_1(t)H(\bm{R})\ket{\psi^R_+(\bm{R})}e^{i\gamma_+(\bm{R})}e^{i\phi_+(t)}
		+C_2(t)H(\bm{R})\ket{\psi^R_-(\bm{R})}e^{i\gamma_-(\bm{R})}e^{i\phi_-(t)}.
	\end{eqnarray}
Multiply both sides of Eq. (\ref{Eq1}) by $\bra{\psi_+^L(\bm{R})}$, we have
\begin{eqnarray}\label{Eq2}
		i\dot{C_1}(t)
		&=&-iC_1(t)\frac{\bra{\psi_+^L(\bm{R})}
			\frac{d\ket{\psi_+^R(\bm{R})}}{d\bm{R}}}
		{\braket{\psi_+^L(\bm{R})|\psi_+^R(\bm{R})}}\frac{d\bm{R}}{dt}
		+C_1(t)\frac{d\gamma_+(\bm{R})}{d\bm{R}}\frac{d\bm{R}}{dt}\nonumber\\
		&&-iC_2(t)\frac{\bra{\psi_+^L(\bm{R})}\frac{d\ket{\psi_-^R(\bm{R})}}{d\bm{R}}}
		{\braket{\psi_+^L(\bm{R})|\psi_+^R(\bm{R})}}\frac{d\bm{R}}{dt}
		e^{i\gamma_-(\bm{R})-i\gamma_+(\bm{R})}e^{i\phi_-(t)-i\phi_+(t)}\nonumber\\
		&&+C_1(t)\dot{\phi_+}(t)
		+C_1(t)\frac{\bra{\psi_+^L(\bm{R})}H(\bm{R})\ket{\psi_+^R(\bm{R})}}
		{\braket{\psi_+^L(\bm{R})|\psi_+^R(\bm{R})}}.
	\end{eqnarray}
Similarly, multiplying both sides of this equation by $\bra{\psi_-^L(\bm{R})}$ yields the following equation for $C_2(t)$:
\begin{eqnarray}\label{Eq3}
		i\dot{C_2}(t)
		&=&-iC_2(t)\frac{\bra{\psi_-^L(\bm{R})}
			\frac{d\ket{\psi_-^R(\bm{R})}}{d\bm{R}}}
		{\braket{\psi_-^L(\bm{R})|\psi_-^R(\bm{R})}}\frac{d\bm{R}}{dt}
		+C_2(t)\frac{d\gamma_-(\bm{R})}{d\bm{R}}\frac{d\bm{R}}{dt}\nonumber\\
		&&-iC_1(t)\frac{\bra{\psi_-^L(\bm{R})}\frac{d\ket{\psi_+^R(\bm{R})}}{d\bm{R}}}
		{\braket{\psi_-^L(\bm{R})|\psi_-^R(\bm{R})}}\frac{d\bm{R}}{dt}
		e^{i\gamma_+(\bm{R})-i\gamma_-(\bm{R})}e^{i\phi_+(t)-i\phi_-(t)}\nonumber\\
		&&+C_2(t)\dot{\phi_-}(t)
		+C_2(t)\frac{\bra{\psi_-^L(\bm{R})}H(\bm{R})\ket{\psi_-^R(\bm{R})}}
		{\braket{\psi_-^L(\bm{R})|\psi_-^R(\bm{R})}}.
	\end{eqnarray}
From Eq. (\ref{Eq2}), we can obtain:
\begin{eqnarray}\label{Eq0}
		i\int_{C_1(0)}^{C_1(T)}\frac{1}{C_1(t)}dC_1(t)
		&=&\int_C[-i\frac{\bra{\psi_+^L(\bm{R})}
			\frac{d\ket{\psi_+^R(\bm{R})}}{d\bm{R}}}
		{\braket{\psi_+^L(\bm{R})|\psi_+^R(\bm{R})}}
		+\frac{d\gamma_+(\bm{R})}{d\bm{R}}]d\bm{R}\nonumber\\
		&&+\int_0^T[\dot{\phi_+}(t)
		+\frac{\bra{\psi_+^L(\bm{R})}H(\bm{R})\ket{\psi_+^R(\bm{R})}}
		{\braket{\psi_+^L(\bm{R})|\psi_+^R(\bm{R})}}]dt\nonumber\\
		&&+\int_0^T\frac{C_2(t)}{C_1(t)}\frac{\bra{\psi_+^L(\bm{R})}\frac{d\ket{\psi_-^R(\bm{R})}}{d\bm{R}}}
		{\braket{\psi_+^L(\bm{R})|\psi_+^R(\bm{R})}}\frac{d\bm{R}}{dt}
		e^{i\gamma_-(\bm{R})-i\gamma_+(\bm{R})}e^{i\phi_-(t)-i\phi_+(t)}dt.
	\end{eqnarray}
If the system is initially prepared in the state $\ket{\psi_+^R(\bm{R})}$ and evolves adiabatically, the adiabatic theorem dictates that the deviation between $\ket{\psi(t)}$ and $\ket{\psi^R_+(\bm{R})}e^{i\gamma_+ (\bm{R})}e^{i\phi_+(t)}$ is very small \cite{Muga,Nenciu,Dridi,Wu19}. From this, it can be inferred that $C_1(t)\approx 1$ and $\int_{C_1(0)}^{C_1(T)}\frac{1}{C_1(t)}dC_1(t)\approx 0$. Additionally, the adiabatic condition for non-Hermitian systems requires that $\mathrm{Re}(i\phi_-(t)-i\phi_+(t))\leq 0$ \cite{Muga,Nenciu,Dridi,Wu19}. Since $C_2(0)=0$, the adiabatic theorem ensures that $C_2(t)e^{i\gamma_-(\bm{R})-i\gamma_+(\bm{R})}e^{i\phi_-(t)-i\phi_+(t)}$ is a small quantity. Moreover, as $\frac{d\bm{R}}{dt}$ is also a small quantity,  the integral $\int_0^T\frac{C_2(t)}{C_1(t)}\frac{\bra{\psi_+^L(\bm{R})}\frac{d\ket{\psi_-^R(\bm{R})}}{d\bm{R}}}
{\braket{\psi_+^L(\bm{R})|\psi_+^R(\bm{R})}}\frac{d\bm{R}}{dt}
e^{i\gamma_-(\bm{R})-i\gamma_+(\bm{R})}e^{i\phi_-(t)-i\phi_+(t)}dt$ is much smaller than other terms and thus negligible.
Under the aforementioned premise, by segregating the zeroth-order term and the first-order term of the derivative of $\bm{R}$ with respect to $t$ in Eq. (\ref{Eq0}), $\gamma_+(\bm{R})$ and $\phi_+(t)$ can be obtained respectively. Analogously, when starting from the initial state $\ket{\psi_-^R(\bm{R})}$ and following an adiabatic path, we derive $\gamma_-(\bm{R})$ and $\phi_-(t)$ from Eq. (\ref{Eq3}). The results are presented below:
\begin{equation}\label{Eq4}
	\gamma_\pm(\bm{R})
	=\int_Ci\frac{\bra{\psi_\pm^L(\bm{R})}
		\frac{d\ket{\psi_\pm^R(\bm{R})}}{d\bm{R}}}
	{\braket{\psi_\pm^L(\bm{R})|\psi_\pm^R(\bm{R})}}d\bm{R},
\end{equation}
\begin{equation}\label{Eq5}
	\phi_\pm(t)=
	-\int_0^T\frac{\bra{\psi_\pm^L(\bm{R})}H(\bm{R})\ket{\psi_\pm^R(\bm{R})}}
	{\braket{\psi_\pm^L(\bm{R})|\psi_\pm^R(\bm{R})}}dt.
\end{equation}
Eq. (\ref{Eq4}) can be used to obtain the Berry connection of the system:
\begin{equation}\label{Eq6}
	\bm{A}^{LR}_\pm=i\frac{\bra{\psi^L_\pm(\bm{R})}\bigtriangledown_{\bm{R}}\ket{\psi^R_\pm(\bm{R})}}
	{\braket{\psi^L_\pm(\bm{R})|\psi^R_\pm(\bm{R})}}.
\end{equation}
Eq. (\ref{Eq6}) involves both left and right eigenstates \cite{Garrison}. The form is indeed simple due to the bi-orthogonality of states. Quantized indices of complex bands are typically obtained using bi-orthogonal relations \cite{rmp1,Garrison,Ueda,NOkuma,np1,Herm2}.

However, we believe that the magnetic field for linear bands can also be obtained without bi-orthogonality. Therefore, we will derive Berry connection represented only by right eigenstates.
If multiply both sides of Eq. (\ref{Eq1}) by $\bra{\psi_+^R(\bm{R})}$ or $\bra{\psi_-^R(\bm{R})}$, the following two equations can be obtained:
\begin{eqnarray}\label{Eq7}
		i\dot{C_1}(t)
		&=&-iC_1(t)\frac{\bra{\psi_+^R(\bm{R})}
			\frac{d\ket{\psi_+^R(\bm{R})}}{d\bm{R}}}
		{\braket{\psi_+^R(\bm{R})|\psi_+^R(\bm{R})}}\frac{d\bm{R}}{dt}
		+C_1(t)\frac{d\gamma_+(\bm{R})}{d\bm{R}}\frac{d\bm{R}}{dt}\nonumber\\
		&&-i\dot{C_2}(t)\frac{\braket{\psi_+^R(\bm{R})|\psi_-^R(\bm{R})}}
		{\braket{\psi_+^R(\bm{R})|\psi_+^R(\bm{R})}}
		e^{i\gamma_-(\bm{R})-i\gamma_+(\bm{R})}e^{i\phi_-(t)-i\phi_+(t)}\nonumber\\
		&&-iC_2(t)\frac{\bra{\psi_+^R(\bm{R})}\frac{d\ket{\psi_-^R(\bm{R})}}{d\bm{R}}}
		{\braket{\psi_+^R(\bm{R})|\psi_+^R(\bm{R})}}\frac{d\bm{R}}{dt}
		e^{i\gamma_-(\bm{R})-i\gamma_+(\bm{R})}e^{i\phi_-(t)-i\phi_+(t)}\nonumber\\
		&&+C_2(t)\frac{\braket{\psi_+^R(\bm{R})|\psi_-^R(\bm{R})}}
		{\braket{\psi_+^R(\bm{R})|\psi_+^R(\bm{R})}}
		\frac{d\gamma_-(\bm{R})}{d\bm{R}}\frac{d\bm{R}}{dt}
		e^{i\gamma_-(\bm{R})-i\gamma_+(\bm{R})}e^{i\phi_-(t)-i\phi_+(t)}\nonumber\\
		&&+C_1(t)\dot{\phi_+}(t)
		+C_2(t)\frac{\braket{\psi_+^R(\bm{R})|\psi_-^R(\bm{R})}}
		{\braket{\psi_+^R(\bm{R})|\psi_+^R(\bm{R})}}\dot{\phi_-}(t)
		e^{i\gamma_-(\bm{R})-i\gamma_+(\bm{R})}e^{i\phi_-(t)-i\phi_+(t)}\nonumber\\
		&&+C_1(t)\frac{\bra{\psi_+^R(\bm{R})}H(\bm{R})\ket{\psi_+^R(\bm{R})}}
		{\braket{\psi_+^R(\bm{R})|\psi_+^R(\bm{R})}}
		+C_2(t)\frac{\bra{\psi_+^R(\bm{R})}H(\bm{R})\ket{\psi_-^R(\bm{R})}}
		{\braket{\psi_+^R(\bm{R})|\psi_+^R(\bm{R})}}
		e^{i\gamma_-(\bm{R})-i\gamma_+(\bm{R})}e^{i\phi_-(t)-i\phi_+(t)},
	\end{eqnarray}
\begin{eqnarray}\label{Eq8}
		i\dot{C_2}(t)
		&=&-iC_2(t)\frac{\bra{\psi_-^R(\bm{R})}
			\frac{d\ket{\psi_-^R(\bm{R})}}{d\bm{R}}}
		{\braket{\psi_-^R(\bm{R})|\psi_-^R(\bm{R})}}\frac{d\bm{R}}{dt}
		+C_2(t)\frac{d\gamma_-(\bm{R})}{d\bm{R}}\frac{d\bm{R}}{dt}\nonumber\\
		&&-i\dot{C_1}(t)\frac{\braket{\psi_-^R(\bm{R})|\psi_+^R(\bm{R})}}
		{\braket{\psi_-^R(\bm{R})|\psi_-^R(\bm{R})}}
		e^{i\gamma_+(\bm{R})-i\gamma_-(\bm{R})}e^{i\phi_+(t)-i\phi_-(t)}\nonumber\\
		&&-iC_1(t)\frac{\bra{\psi_-^R(\bm{R})}\frac{d\ket{\psi_+^R(\bm{R})}}{d\bm{R}}}
		{\braket{\psi_-^R(\bm{R})|\psi_-^R(\bm{R})}}\frac{d\bm{R}}{dt}
		e^{i\gamma_+(\bm{R})-i\gamma_-(\bm{R})}e^{i\phi_+(t)-i\phi_-(t)}\nonumber\\
		&&+C_1(t)\frac{\braket{\psi_-^R(\bm{R})|\psi_+^R(\bm{R})}}
		{\braket{\psi_-^R(\bm{R})|\psi_-^R(\bm{R})}}
		\frac{d\gamma_+(\bm{R})}{d\bm{R}}\frac{d\bm{R}}{dt}
		e^{i\gamma_+(\bm{R})-i\gamma_-(\bm{R})}e^{i\phi_+(t)-i\phi_-(t)}\nonumber\\
		&&+C_2(t)\dot{\phi_-}(t)
		+C_1(t)\frac{\braket{\psi_-^R(\bm{R})|\psi_+^R(\bm{R})}}
		{\braket{\psi_-^R(\bm{R})|\psi_-^R(\bm{R})}}\dot{\phi_+}(t)
		e^{i\gamma_+(\bm{R})-i\gamma_-(\bm{R})}e^{i\phi_+(t)-i\phi_-(t)}\nonumber\\
		&&+C_2(t)\frac{\bra{\psi_-^R(\bm{R})}H(\bm{R})\ket{\psi_-^R(\bm{R})}}
		{\braket{\psi_-^R(\bm{R})|\psi_-^R(\bm{R})}}
		+C_1(t)\frac{\bra{\psi_-^R(\bm{R})}H(\bm{R})\ket{\psi_+^R(\bm{R})}}
		{\braket{\psi_-^R(\bm{R})|\psi_-^R(\bm{R})}}
		e^{i\gamma_+(\bm{R})-i\gamma_-(\bm{R})}e^{i\phi_+(t)-i\phi_-(t)}.
	\end{eqnarray}
Substitute Eq. (\ref{Eq8}) into Eq. (\ref{Eq7}) to eliminate $i\dot{C_2}(t)$, and the same operation to eliminate $i\dot{C_1}(t)$ from Eq. (\ref{Eq8}), Eq. (\ref{Eq7}) and (\ref{Eq8}) can be further simplified as follows:
\begin{eqnarray}\label{Eq9}
		&&i\dot{C_1}(t)\nonumber\\
		&&=C_1(t)\frac{d\gamma_+(\bm{R})}{d\bm{R}}\frac{d\bm{R}}{dt}\nonumber\\
		&&-iC_1(t)\frac{\braket{\psi_-^R(\bm{R})|\psi_-^R(\bm{R})}
			\bra{\psi_+^R(\bm{R})}\frac{d\ket{\psi_+^R(\bm{R})}}{d\bm{R}}\frac{d\bm{R}}{dt}
			-\braket{\psi_+^R(\bm{R})|\psi_-^R(\bm{R})}
			\bra{\psi_-^R(\bm{R})}\frac{d\ket{\psi_+^R(\bm{R})}}{d\bm{R}}\frac{d\bm{R}}{dt}}
		{\braket{\psi_+^R(\bm{R})|\psi_+^R(\bm{R})}
			\braket{\psi_-^R(\bm{R})|\psi_-^R(\bm{R})}
			-\braket{\psi_+^R(\bm{R})|\psi_-^R(\bm{R})}
			\braket{\psi_-^R(\bm{R})|\psi_+^R(\bm{R})}}\nonumber\\
		&&-iC_2(t)\frac{\braket{\psi_-^R(\bm{R})|\psi_-^R(\bm{R})}
			\bra{\psi_+^R(\bm{R})}\frac{d\ket{\psi_-^R(\bm{R})}}{d\bm{R}}\frac{d\bm{R}}{dt}
			-\braket{\psi_+^R(\bm{R})|\psi_-^R(\bm{R})}
			\bra{\psi_-^R(\bm{R})}\frac{d\ket{\psi_-^R(\bm{R})}}{d\bm{R}}\frac{d\bm{R}}{dt}}
		{\braket{\psi_+^R(\bm{R})|\psi_+^R(\bm{R})}
			\braket{\psi_-^R(\bm{R})|\psi_-^R(\bm{R})}
			-\braket{\psi_+^R(\bm{R})|\psi_-^R(\bm{R})}
			\braket{\psi_-^R(\bm{R})|\psi_+^R(\bm{R})}}
		e^{i\gamma_-(\bm{R})-i\gamma_+(\bm{R})}e^{i\phi_-(t)-i\phi_+(t)}\nonumber\\
		&&+C_1(t)\dot{\phi_+}(t)\nonumber\\
		&&+C_1(t)\frac{\braket{\psi_-^R(\bm{R})|\psi_-^R(\bm{R})}
			\bra{\psi_+^R(\bm{R})}H(\bm{R})\ket{\psi_+^R(\bm{R})}
			-\braket{\psi_+^R(\bm{R})|\psi_-^R(\bm{R})}
			\bra{\psi_-^R(\bm{R})}H(\bm{R})\ket{\psi_+^R(\bm{R})}}
		{\braket{\psi_+^R(\bm{R})|\psi_+^R(\bm{R})}
			\braket{\psi_-^R(\bm{R})|\psi_-^R(\bm{R})}
			-\braket{\psi_+^R(\bm{R})|\psi_-^R(\bm{R})}
			\braket{\psi_-^R(\bm{R})|\psi_+^R(\bm{R})}},
	\end{eqnarray}

\begin{eqnarray}\label{Eq10}
	&&i\dot{C_2}(t) \nonumber\\
		&&=C_2(t)\frac{d\gamma_-(\bm{R})}{d\bm{R}}\frac{d\bm{R}}{dt} \nonumber\\
		&&-iC_2(t)\frac{\braket{\psi_+^R(\bm{R})|\psi_+^R(\bm{R})}
			\bra{\psi_-^R(\bm{R})}\frac{d\ket{\psi_-^R(\bm{R})}}{d\bm{R}}\frac{d\bm{R}}{dt}
			-\braket{\psi_-^R(\bm{R})|\psi_+^R(\bm{R})}
			\bra{\psi_+^R(\bm{R})}\frac{d\ket{\psi_-^R(\bm{R})}}{d\bm{R}}\frac{d\bm{R}}{dt}}
		{\braket{\psi_-^R(\bm{R})|\psi_-^R(\bm{R})}
			\braket{\psi_+^R(\bm{R})|\psi_+^R(\bm{R})}
			-\braket{\psi_-^R(\bm{R})|\psi_+^R(\bm{R})}
			\braket{\psi_+^R(\bm{R})|\psi_-^R(\bm{R})}}\nonumber\\
		&&-iC_1(t)\frac{\braket{\psi_+^R(\bm{R})|\psi_+^R(\bm{R})}
			\bra{\psi_-^R(\bm{R})}\frac{d\ket{\psi_+^R(\bm{R})}}{d\bm{R}}\frac{d\bm{R}}{dt}
			-\braket{\psi_-^R(\bm{R})|\psi_+^R(\bm{R})}
			\bra{\psi_+^R(\bm{R})}\frac{d\ket{\psi_+^R(\bm{R})}}{d\bm{R}}\frac{d\bm{R}}{dt}}
		{\braket{\psi_-^R(\bm{R})|\psi_-^R(\bm{R})}
			\braket{\psi_+^R(\bm{R})|\psi_+^R(\bm{R})}
			-\braket{\psi_-^R(\bm{R})|\psi_+^R(\bm{R})}
			\braket{\psi_+^R(\bm{R})|\psi_-^R(\bm{R})}}
		e^{i\gamma_+(\bm{R})-i\gamma_-(\bm{R})}e^{i\phi_+(t)-i\phi_-(t)}\nonumber\\
		&&+C_2(t)\dot{\phi_-}(t)\nonumber\\
		&&+C_2(t)\frac{\braket{\psi_+^R(\bm{R})|\psi_+^R(\bm{R})}
			\bra{\psi_-^R(\bm{R})}H(\bm{R})\ket{\psi_-^R(\bm{R})}
			-\braket{\psi_-^R(\bm{R})|\psi_+^R(\bm{R})}
			\bra{\psi_+^R(\bm{R})}H(\bm{R})\ket{\psi_-^R(\bm{R})}}
		{\braket{\psi_-^R(\bm{R})|\psi_-^R(\bm{R})}
			\braket{\psi_+^R(\bm{R})|\psi_+^R(\bm{R})}
			-\braket{\psi_-^R(\bm{R})|\psi_+^R(\bm{R})}
			\braket{\psi_+^R(\bm{R})|\psi_-^R(\bm{R})}}.
	\end{eqnarray}
When the system undergoes adiabatic evolution starting from the initial state $\ket{\psi_+^R(\bm{R})}$, via the same analysis as for Eq. (\ref{Eq2}), we derive the expressions for $\gamma_+$ and $\phi_+(t)$ from Eq. (\ref{Eq9}). Analogously, by applying the same methodology to Eq. (\ref{Eq10}) we obtain the corresponding expressions for $\gamma_-(\bm{R})$ and $\phi_-(t)$. The results are presented below:
\begin{equation}\label{Eq11}
	\gamma_\pm(\bm{R})
	=\int_Ci\frac{\braket{\psi_\mp^R(\bm{R})|\psi_\mp^R(\bm{R})}
		\bra{\psi_\pm^R(\bm{R})}\frac{d\ket{\psi_\pm^R(\bm{R})}}{d\bm{R}}
		-\braket{\psi_\pm^R(\bm{R})|\psi_\mp^R(\bm{R})}
		\bra{\psi_\mp^R(\bm{R})}\frac{d\ket{\psi_\pm^R(\bm{R})}}{d\bm{R}}}
	{\braket{\psi_\pm^R(\bm{R})|\psi_\pm^R(\bm{R})}
		\braket{\psi_\mp^R(\bm{R})|\psi_\mp^R(\bm{R})}
		-\braket{\psi_\pm^R(\bm{R})|\psi_\mp^R(\bm{R})}
		\braket{\psi_\mp^R(\bm{R})|\psi_\pm^R(\bm{R})}}d\bm{R},
\end{equation}
\begin{equation}\label{Eq12}
	\phi_\pm(t)=
	-\int_0^T\frac{\braket{\psi_\mp^R(\bm{R})|\psi_\mp^R(\bm{R})}
		\bra{\psi_\pm^R(\bm{R})}H(\bm{R})\ket{\psi_\pm^R(\bm{R})}
		-\braket{\psi_\pm^R(\bm{R})|\psi_\mp^R(\bm{R})}
		\bra{\psi_\mp^R(\bm{R})}H(\bm{R})\ket{\psi_\pm^R(\bm{R})}}
	{\braket{\psi_\pm^R(\bm{R})|\psi_\pm^R(\bm{R})}
		\braket{\psi_\mp^R(\bm{R})|\psi_\mp^R(\bm{R})}
		-\braket{\psi_\pm^R(\bm{R})|\psi_\mp^R(\bm{R})}
		\braket{\psi_\mp^R(\bm{R})|\psi_\pm^R(\bm{R})}}dt.
\end{equation}
Eq. (\ref{Eq11}) can be used to obtain the Berry connection of the system:
\begin{equation}\label{Eq13}
	\bm{\tilde{A}}^{RR}_\pm=i\frac{\braket{\psi_\mp^R(\bm{R})|\psi_\mp^R(\bm{R})}
		\bra{\psi_\pm^R(\bm{R})}\bigtriangledown_{\bm{R}}\ket{\psi_\pm^R(\bm{R})}
		-\braket{\psi_\pm^R(\bm{R})|\psi_\mp^R(\bm{R})}
		\bra{\psi_\mp^R(\bm{R})}\bigtriangledown_{\bm{R}}\ket{\psi_\pm^R(\bm{R})}}
	{\braket{\psi_\pm^R(\bm{R})|\psi_\pm^R(\bm{R})}
		\braket{\psi_\mp^R(\bm{R})|\psi_\mp^R(\bm{R})}
		-\braket{\psi_\pm^R(\bm{R})|\psi_\mp^R(\bm{R})}
		\braket{\psi_\mp^R(\bm{R})|\psi_\pm^R(\bm{R})}}.
\end{equation}
When the adiabatic conditions are satisfied, Eq. (\ref{Eq13}) is equivalent to Eq. (\ref{Eq6}).

Additionally, another form of connection represented by the right vector is given by \begin{equation}
\bm{A}^{RR}_\pm=i\frac{\bra{\psi^R_\pm(\bm{R})}\bigtriangledown_{\bm{R}}\ket{\psi^R_\pm(\bm{R})}}
{\braket{\psi^R_\pm(\bm{R})|\psi^R_\pm(\bm{R})}}.
\end{equation}
This expression cannot be derived from the above fundamental principles, but it is given by the analogy with the Berry connection in Hermitian systems. For Hermitian systems, $\bm{\tilde{A}}^{RR}_\pm=\bm{A}^{RR}_\pm$. However, for non-Hermitian systems, the two connections are not equal due to the non-orthogonality of eigenstates. The Berry curvature (effective magnetic field) in parameter space is obtained by taking the curl of the Berry connection. In non-Hermitian systems, the magnetic fields $\bm{\tilde{B}}^{RR}_\pm$ and $\bm{B}^{RR}_\pm$ are generally distinct, even though their total magnetic charges are identical \cite{Shen,Silberstein,Nakamura,Jezequel}. This implies that different geometric phases can be derived from the two connections, where
\begin{eqnarray}
	\oint_C \bm{\tilde{A}}^{RR}_\pm \cdot d \bm{R}= \oint_C \bm{A}^{LR}_\pm \cdot d \bm{R} , \label{vt0}	
\end{eqnarray}
\begin{eqnarray}
	\oint_C \bm{A}^{RR}_\pm \cdot d \bm{R}. \label{motion0}
\end{eqnarray}
To explicitly illustrate the differences and conveniently test the reasonableness of each form, we consider a two-level non-Hermitian system described by the following Hamiltonian:
\begin{equation}\label{Hamiltonian}
	H(\bm{R})=
	\begin{pmatrix}
		Z+iZ_0&X-iY\\
		X+iY&-Z-iZ_0
	\end{pmatrix},
\end{equation}
where $\bm{R}=(X,Y,Z)$, $Z_0$ describes the non-Hermitian effects. The eigenvalue of Eq. (\ref{Hamiltonian}) is $E_\pm(\bm{R})=\pm\sqrt{X^2+Y^2+Z^2-Z_0^2+2iZZ_0}$.
For such a non-Hermitian system, its parameter space forms a Riemannian surface \cite{Wu,IIArkhipov,KWang,HNasari}, and different methods of cutting this Riemannian surface result in distinct selections of eigenstates \cite{Wu}.
From the perspective of physical measurements, it is considered more appropriate to mark the different eigenvalues by the real part of the energy spectrum \cite{QZhong}, as their differences are easily measurable via optical spectroscopic measurements.
Based on this, we express $E$ as $E_\pm=\pm(a+ib)$ and select branch by setting $a>0$ and $\sqrt{(X^2+Y^2+Z^2-Z_0^2)^2+4(ZZ_0)^2}>0$. This yields: $a=\sqrt{\frac{X^2+Y^2+Z^2-Z_0^2+\sqrt{(X^2+Y^2+Z^2-Z_0^2)^2+4(ZZ_0)^2}}{2}}$, $b=\frac{ZZ_0}{a}$.
The eigenstates of the system can be expressed as follows:
\begin{equation}\label{Eq15}
	\ket{\psi^R_\pm(\bm{R})}=
	\begin{pmatrix}
		X-iY\\
		-Z-iZ_0+E_\pm
	\end{pmatrix}.
\end{equation}
Taking the state $\ket{\psi^R_+(\bm{R})}$ as an example, calculating the curl of the connection yields two distinct magnetic fields:
\begin{eqnarray}\label{Eq22}
		\bm{B}^{RR}_+&=&
		\frac{2(X^2+Y^2)(YZ_0-aX)+2[X^2+Y^2+(a-Z)^2+(b-Z_0)^2](XZ-YZ_0)}
		{(a^2+b^2)[X^2+Y^2+(a-Z)^2+(b-Z_0)^2]^2}\bm{e_X}\nonumber\\
		&&+\frac{2(X^2+Y^2)(-XZ_0-aY)+2[X^2+Y^2+(a-Z)^2+(b-Z_0)^2](YZ+XZ_0)}
		{(a^2+b^2)[X^2+Y^2+(a-Z)^2+(b-Z_0)^2]^2}\bm{e_Y}\nonumber\\
		&&+\frac{-b}{2Z_0(a^2+b^2)}\bm{e_Z},
	\end{eqnarray}
\begin{equation}
	\bm{\tilde{B}}^{RR}_+=\bm{B}^{LR}_+=Re(\bm{B}^{LR}_+)+iIm(\bm{B}^{LR}_+),
\end{equation}
where
\begin{eqnarray}\label{Eq20}
		Re(\bm{B}^{LR}_+)
		&=&\frac{-X(a^3-3ab^2)}{2[(a^3-3ab^2)^2+(3a^2b-b^3)^2]}\bm{e_X}+\frac{-Y(a^3-3ab^2)}{2[(a^3-3ab^2)^2+(3a^2b-b^3)^2]}\bm{e_Y}\nonumber\\
		&&+\frac{-Z(a^3-3ab^2)-Z_0(3a^2b-b^3)}{2[(a^3-3ab^2)^2+(3a^2b-b^3)^2])}\bm{e_Z},
	\end{eqnarray}
\begin{eqnarray}\label{Eq21}
		Im(\bm{B}^{LR}_+)
		&=&\frac{X(3a^2b-b^3)}{2[(a^3-3ab^2)^2+(3a^2b-b^3)^2]}\bm{e_X}+\frac{Y(3a^2b-b^3)}{2[(a^3-3ab^2)^2+(3a^2b-b^3)^2]}\bm{e_Y}\nonumber\\
		&&+\frac{Z(3a^2b-b^3)-Z_0(a^3-3ab^2)}{2[(a^3-3ab^2)^2+(3a^2b-b^3)^2])}\bm{e_Z}.
	\end{eqnarray}
A distinct difference is observed between $\bm{\tilde{B}}_+^{RR}$ and $\bm{B}^{RR}_+$. Specifically, $\bm{\tilde{B}}^{RR}_+$ corresponds to a complex magnetic field, whereas $\bm{B}^{RR}_+$ represents a real magnetic field. As the system evolves along a closed trajectory in the parameter space, different magnetic fields give rise to distinct geometric phases. Accordingly, it is imperative to investigate the correspondence between these two types of magnetic fields and real-world scenarios via numerical experiments.

\section*{Appendix B: A striking relation $\Delta \phi_d=\Delta \phi_g$}
To confirm which Berry connection is reasonable for calculating geometric phases, we performed a numerical experiment to observe the phase evolution directly.
The initial state was selected as $\ket{\psi^R_+(\bm{R})}$.
From the expression of the system energy, it can be observed that $b>0$, when $Z>0$. In this case, the amplitude of $\ket{\psi^R_+(\bm{R})}e^{i\phi_+(t)}$ over the course of evolution, while that of $\ket{\psi^R_-(\bm{R})}e^{i\phi_-(t)}$ decreases. By the adiabatic theorem for non-Hermitian systems \cite{Muga,Nenciu,Dridi,Wu19}, we thus infer that in the region $Z>0$, state $\ket{\psi^R_+(\bm{R})}$ can remain adiabatic.
Based on the above analysis, the evolution was conducted at the parameter space position $Z=0.5$ to ensure adiabaticity of the system. We set $X=r\cos(\omega t)$, $Y=r\sin(\omega t)$ ($r=\sqrt{X^2+Y^2}$), and the evolution trajectory is: $\omega t: 0\to 2\pi$. Moreover, the adiabatic theorem requires the system to evolve sufficiently slowly, which we ensure by choosing a small value of $\omega$ ($\omega=0.0005\pi$). Under such conditions, the fidelity $f_+(t)=|\braket{\psi(t)|\psi_+^R(\bm{R})}|^2$ remains very close to 1 at all times during evolution.

In general, the geometric phase $\phi_g$ is given by the difference between the total phase $\phi_{total}$ and the dynamic phase $\phi_{d}$: $\phi_g=\phi_{total}-\phi_d$.
Notably, the dynamical phase can be calculated using two distinct methods, denoted as $\phi_{d}$ and $\tilde{\phi}_{d}$, respectively.
The first form for the geometric phase can be described by
\begin{eqnarray}\label{gphase1}
	\phi_g&=&\phi_{total}-\phi_{d}\nonumber\\
	&=& \phi_{total}-[-\int_{0}^{T}E(t)dt],
\end{eqnarray}
where
$
	E(t)=\frac{\bra{\psi(t)}H\ket{\psi(t)}}{\braket{\psi(t)|\psi(t)}}
$ denotes the time-dependent energy expectation value.
The second form for the geometric phase is
\begin{eqnarray}\label{gphase2}
	\tilde{\phi}_g &=&\phi_{total}-\tilde{\phi}_{d} \nonumber\\
	&=& \phi_{total}-[-\int_{0}^{T}E_+(\bm{R})dt] ,
\end{eqnarray}
where $E_+(\bm{R})$ is the instantaneous eigenvalue from
$H(\bm{R}) \ket{\psi^R_+(\bm{R})} =\ E_+(\bm{R})  \ket{\psi^R_+(\bm{R})}.
$

\begin{figure}[H]
	\centering
	\includegraphics[width=17.5cm,height=3.5cm]{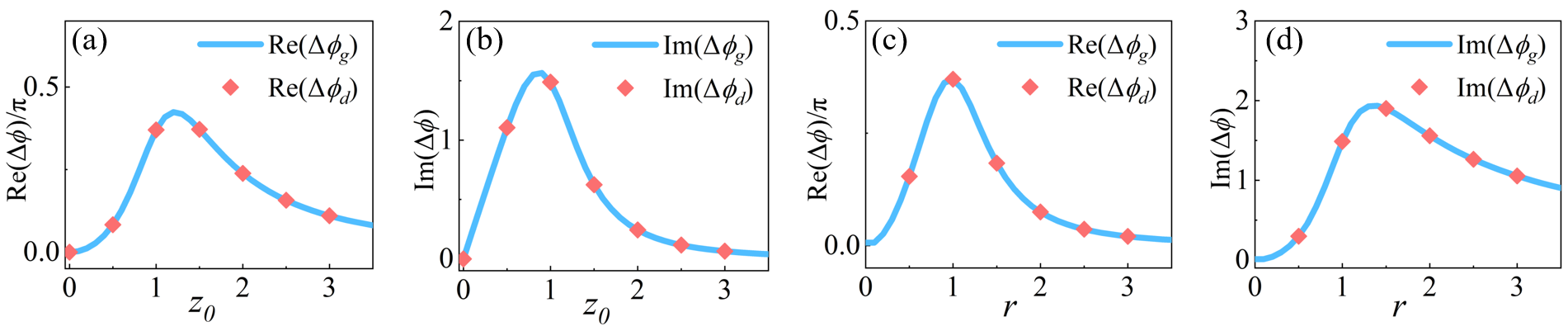}
	\caption{\label{4}(a) $\mathrm{Re}(\Delta \phi_g)$ and $\mathrm{Re}(\Delta \phi_d)$ at different $Z_0$, when $Z=0$, $r=1$, $\omega=0.0005\pi$. (b) $\mathrm{Im}(\Delta \phi_g)$ and $\mathrm{Im}(\Delta \phi_d)$ at different $z_0$, when $Z=0$, $r=1$, $\omega=0.0005\pi$. (c) $\mathrm{Re}(\Delta \phi_g)$ and $\mathrm{Re}(\Delta \phi_d)$ at different $r$, when $Z=0$, $Z_0=1$, $\omega=0.0005\pi$. (d) $\mathrm{Im}(\Delta \phi_g)$ and $\mathrm{Im}(\Delta \phi_d)$ at different $r$, when $Z=0$, $Z_0=1$, $\omega=0.0005\pi$.}
\end{figure}
It should be noted that due to the extremely small value of $\omega$, the amplitude of the state may grow excessively during numerical evolution, exceeding the computational range of the system. To mitigate this issue, a compensatory phase factor $e^{iE(t)}$ (or $e^{iE_+(\bm{R})}$) is applied at each time step to cancel the contribution of the dynamical phase. After completing one full cycle of evolution, the geometric phase $\phi_g$ (or $\tilde{\phi}_g$)can be directly obtained by measuring the phase difference between the initial and final states. This operation is performed solely for the convenience of numerical computation and, within the selected model, does not affect the measured geometric phase.

Based on $\tilde{\bm{A}}_+^{RR}$ and $\bm{A}_+^{RR}$, the theoretical geometric phases can be calculated respectively as $\tilde{\phi}_g^{RR} = \oint_C \bm{\tilde{A}}_+^{RR} \cdot d \bm{l}= \oint_C \bm{A}_+^{LR} \cdot d \bm{l}$ and $\phi_g^{RR} = \oint_C \bm{A}_+^{RR} \cdot d \bm{l}$. A comparison between the theoretically derived and numerically computed phases reveals $ \tilde{\phi}_g=\tilde{\phi}_g^{RR}$ and $\phi_g=\phi_g^{RR}$. Then we obtain
\begin{eqnarray}\label{deltad0}
	\Delta \phi_d &=& \Delta \phi_g  \nonumber\\
	&=&  \int_{0}^{T}\Delta E(t) dt,
\end{eqnarray}
where $\Delta \phi_g=\tilde{\phi_g}^{RR}-\phi_g^{RR}$, $\phi_{d} -\tilde{\phi}_d=\Delta \phi_d$, and $\Delta E(t)=E_+(\bm{R})-E(t)$. We uncover a general relationship between the accumulation of small energy differences ($\Delta \phi_d$) and the difference in geometric phases ($\Delta \phi_g$) derived from the distinct Berry connection forms ($\bm{\tilde{A}}_+^{RR}$ and $\bm{A}_+^{RR}$).

This result is more intuitively illustrated in Fig. \ref{4}. Fig. \ref{4} (a) and (b) show the two phase differences with $Z_0$ under the conditions $Z=0.5$ and $r=1$. It is seen that the two phase differences are equal. When $Z_0=0$, the system reduces to a Hermitian system, and both phase differences vanish. By fixing $Z_0=1$ and $Z=0.5$ while varying $r$, we obtain Fig. \ref{4} (c) and (d), where the curves are also found to satisfy Eq. (\ref{deltad0}).

The above relation means that the phase $\Delta \phi_d$ is of geometric rather than dynamical nature. We check this by observing $\Delta \phi_d$ with varied small changing rates of $\omega$.  With $\omega$ kept extremely small, varying $\omega$ does not result in significant changes in the value of $\Delta \phi_d$, as shown in Fig. \ref{5} (a). In Fig. \ref{5} (b), we display the evolution of $\Delta E(t)$ for different values of $\omega$ throughout the process. Combining the insights from Fig. \ref{5} (a), (b)and (c), it can be observed that although the value of $\Delta E(t)$ at each moment differs for different $\omega$, the total cumulative amount after evolution remains unchanged. These indicate that when the system effectively maintains adiabaticity, the value of $\Delta \phi_d$ remains largely invariant over the evolution time and thus it indeed is geometric.

\begin{figure}[H]
	\centering
	\includegraphics[width=14.9cm,height=4.0cm]{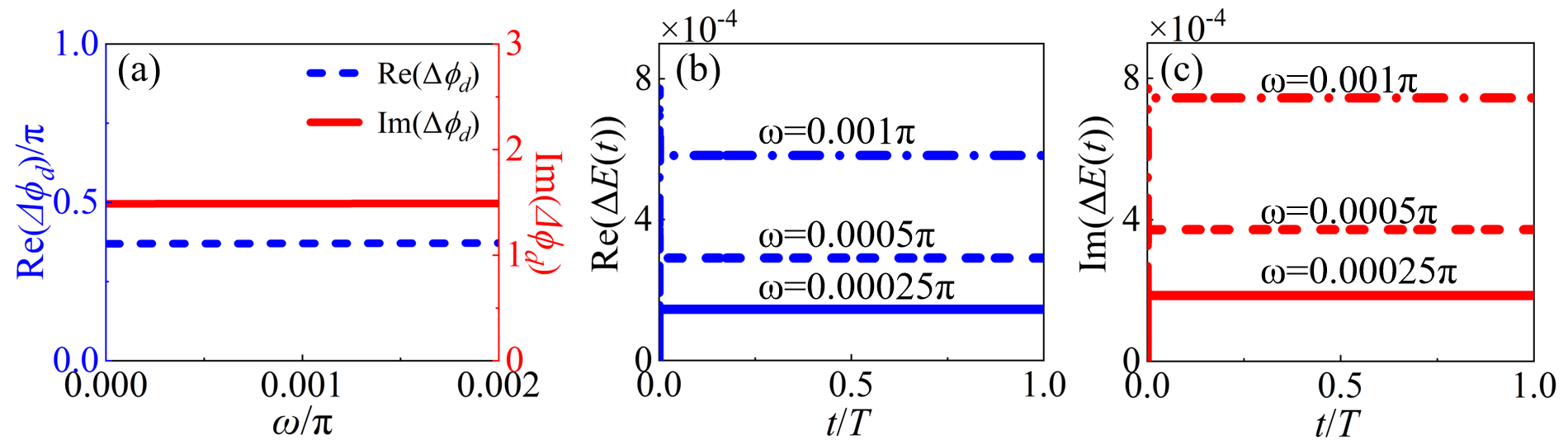}
	\caption{\label{5}(a) $\mathrm{Re}(\Delta \phi_d)$ and $\mathrm{Im}(\Delta \phi_d)$ at different $\omega$, when $Z=0.5$, $r=1$, $Z_0=1$. (b) The change of $\mathrm{Re}(\Delta E(t))$ during evolution at different $\omega$. (c) The change of $\mathrm{Im}(\Delta E(t))$ during evolution at different $\omega$.}
\end{figure}

\section*{Appendix C: Proof of $\Delta \phi_g\approx \Delta \phi_d$ under adiabatic conditions}
We proceed to derive analytically Eq. (\ref{deltad0}) with the adiabatic conditions holding. According to expansion $\ket{\psi(t)}=C_1(t)\ket{\psi^R_+(\bm{R})}e^{i\gamma_+ (\bm{R})}e^{i\phi_+(t)}+C_2(t)\ket{\psi^R_-(\bm{R})}e^{i\gamma_- (\bm{R})}e^{i\phi_-(t)}$, the energy expectation of the system at each moment can be written as:
\begin{equation}\label{Et}
	\begin{aligned}
&E(t)=\frac{\bra{\psi(t)}H\ket{\psi(t)}}{\braket{\psi(t)|\psi(t)}}\\
=&\frac{[C_1^*(t)\bra{\psi_+^R(\bm{R})}
	+C_2^*(t)\bra{\psi_-^R(\bm{R})}e^{-i\gamma_-^*+i\gamma_+^*}
	e^{-i\phi_-^*+i\phi_+^*}]
	H(\bm{R})
	[C_1(t)\ket{\psi^R_+(\bm{R})}+C_2(t)\ket{\psi^R_-(\bm{R})}e^{i\gamma_- -i\gamma_+}e^{i\phi_- -i\phi_+}]}
{[C_1^*(t)\bra{\psi_+^R(\bm{R})}
		+C_2^*(t)\bra{\psi_-^R(\bm{R})}e^{-i\gamma_-^*+i\gamma_+^*}e^{-i\phi_-^*+i\phi_+^*}]
		[C_1(t)\ket{\psi^R_+(\bm{R})}+C_2(t)\ket{\psi^R_-(\bm{R})}e^{i\gamma_- -i\gamma_+}e^{i\phi_- -i\phi_+}]},
	\end{aligned}
\end{equation}
where $\gamma_\pm (\bm{R})$ is abbreviated as $\gamma_\pm$ and $\phi_\pm(t)$ is abbreviated as $\phi_\pm$. Taking $\ket{\psi^R_+(\bm{R})}$ as the initial state, when the system remains adiabatic, $C_2(t)e^{i\gamma_- -i\gamma_+}e^{i\phi_- -i\phi_+}$ is a small quantity. Therefore, we ignore the higher-order terms of $C_2(t)e^{i\gamma_- -i\gamma_+}e^{i\phi_- -i\phi_+}$ and obtain:
\begin{equation}\label{Et1}
	\begin{aligned}
&E(t)\\%
\approx  &\frac{|C_1(t)|^2E_+(\bm{R})\braket{\psi_+^R(\bm{R})|\psi_+^R(\bm{R})}+
|C_1(t)||C_{02}(t)|
[E_-(\bm{R})\braket{\psi_+^R(\bm{R})|\psi_-^R(\bm{R})}e^{i\beta-i\alpha}
+E_+(\bm{R})\braket{\psi_-^R(\bm{R})|\psi_+^R(\bm{R})}e^{-i\beta+i\alpha}]}
{|C_1(t)|^2\braket{\psi_+^R(\bm{R})|\psi_+^R(\bm{R})}+
	|C_1(t)||C_{02}(t)|
	[\braket{\psi_+^R(\bm{R})|\psi_-^R(\bm{R})}e^{i\beta-i\alpha}
	+\braket{\psi_-^R(\bm{R})|\psi_+^R(\bm{R})}e^{-i\beta+i\alpha}]},
\end{aligned}
\end{equation}
where $\alpha$ and $\beta$ are the arguments of $C_1(t)$ and $C_{02}(t)$ respectively and $C_{02}(t)=C_2(t)e^{i\gamma_- -i\gamma_+}e^{i\phi_- -i\phi_+}$. $E(t)$ takes the form of $\frac{n+\varepsilon n_1}{m+\varepsilon m_1}$, where $\varepsilon$ represents a small quantity corresponding to $|C_{02}|$ in $E(t)$. Since $\frac{n+\varepsilon n_1}{m+\varepsilon m_1}\approx \frac{n}{m}+\frac{(n_1m-m_1n)\varepsilon}{m^2}+\mathscr{O}(\varepsilon^2)$, we can perform a small $|C_{02}(t)|$ expansion on Eq. (\ref{Et1}), we obtain:
\begin{equation}\label{deltaE}
	\begin{aligned}
E_+(\bm{R})-E(t)\approx [E_+(\bm{R})-E_-(\bm{R})]\frac{\braket{\psi_+^R(\bm{R})|\psi_-^R(\bm{R})}}
{\braket{\psi_+^R(\bm{R})|\psi_+^R(\bm{R})}}\frac{C_2(t)}{C_1(t)}
e^{i\gamma_-(\bm{R}) -i\gamma_+(\bm{R})}e^{i\phi_-(t) -i\phi_+(t)}.
	\end{aligned}
\end{equation}
When the system remains adiabatic, $C_1(t)\approx 1$. Integrating both sides of Eq. (\ref{deltaE}) with respect to $t$ yields:
\begin{equation}\label{intdeltaE}
	\begin{aligned}
\int_0^T[E_+(\bm{R})-E(t)]dt\approx \int_0^TC_2(t)[E_+(\bm{R})-E_-(\bm{R})]\frac{\braket{\psi_+^R(\bm{R})|\psi_-^R(\bm{R})}}
		{\braket{\psi_+^R(\bm{R})|\psi_+^R(\bm{R})}}
		e^{i\gamma_-(\bm{R}) -i\gamma_+(\bm{R})}e^{i\phi_-(t) -i\phi_+(t)}dt.
	\end{aligned}
\end{equation}

Additionally, from Eq. (\ref{Eq5}) and Eq. (\ref{Eq11}), we can obtain $\dot{\phi_+}=-E_+(\bm{R})$,  $\dot{\phi_-}\frac{\braket{\psi_+^R(\bm{R})|\psi_-^R(\bm{R})}}
{\braket{\psi_+^R(\bm{R})|\psi_-^R(\bm{R})}}=\frac{\bra{\psi_+^R(\bm{R})}H(\bm{R})\ket{\psi_-^R(\bm{R})}}
{\braket{\psi_+^R(\bm{R})|\psi_+^R(\bm{R})}}$ and $\frac{d\gamma_+(\bm{R})}{d\bm{R}}\frac{d\bm{R}}{dt}=\tilde{\bm{A}}^{RR}_+$. Substituting these expressions into Eq. (\ref{Eq7}) integrating it yields:
\begin{eqnarray}\label{deltaA1}
		&&\int_C(\tilde{\bm{A}}^{RR}_+-\bm{A}^{RR}_+)d\bm{R}\nonumber\\
		&=&i\int_{C_1(0)}^{C_1(T)}\frac{1}{C_1(t)}dC_1(t)\nonumber\\
		&&+\int_0^T i\frac{\dot{C_2}(t)}{C_1(t)}\frac{\braket{\psi_+^R(\bm{R})|\psi_-^R(\bm{R})}}
		{\braket{\psi_+^R(\bm{R})|\psi_+^R(\bm{R})}}
		e^{i\gamma_-(\bm{R})-i\gamma_+(\bm{R})}e^{i\phi_-(t)-i\phi_+(t)}dt\nonumber\\
		&&+\int_0^T i\frac{C_2(t)}{C_1(t)}\frac{\bra{\psi_+^R(\bm{R})}\frac{d\ket{\psi_-^R(\bm{R})}}{d\bm{R}}}
		{\braket{\psi_+^R(\bm{R})|\psi_+^R(\bm{R})}}\frac{d\bm{R}}{dt}
		e^{i\gamma_-(\bm{R})-i\gamma_+(\bm{R})}e^{i\phi_-(t)-i\phi_+(t)}dt\nonumber\\
		&&-\int_0^T \frac{C_2(t)}{C_1(t)}\frac{\braket{\psi_+^R(\bm{R})|\psi_-^R(\bm{R})}}
		{\braket{\psi_+^R(\bm{R})|\psi_+^R(\bm{R})}}
		\frac{d\gamma_-(\bm{R})}{d\bm{R}}\frac{d\bm{R}}{dt}
		e^{i\gamma_-(\bm{R})-i\gamma_+(\bm{R})}e^{i\phi_-(t)-i\phi_+(t)}dt.
	\end{eqnarray}
When the system satisfies the adiabatic condition, $C_1(t)\approx 1$ and $\int_{C_1(0)}^{C_1(T)}\frac{1}{C_1(t)}dC_1(t)\approx 0$. Since $C_2(t)e^{i\gamma_-(\bm{R})-i\gamma_+(\bm{R})}e^{i\phi_-(t)-i\phi_+(t)}$ is a small quantity, and $\frac{d\bm{R}}{dt}$ also remains small under the adiabatic condition, The last two terms of Eq. (\ref{deltaA1}) can be neglected, so
\begin{eqnarray}\label{intdeltaA1}
		&&\int_C(\tilde{\bm{A}}^{RR}_+-\bm{A}^{RR}_+)d\bm{R}\nonumber\\
		&\approx&
		\int_0^Ti\dot{C_2}(t)\frac{\braket{\psi_+^R(\bm{R})|\psi_-^R(\bm{R})}}
		{\braket{\psi_+^R(\bm{R})|\psi_+^R(\bm{R})}}
		e^{i\gamma_-(\bm{R})-i\gamma_+(\bm{R})}e^{i\phi_-(t)-i\phi_+(t)}dt\nonumber\\
		&=&
		iC_2(t)\frac{\braket{\psi_+^R(\bm{R})|\psi_-^R(\bm{R})}}
		{\braket{\psi_+^R(\bm{R})|\psi_+^R(\bm{R})}}
		e^{i\gamma_-(\bm{R})-i\gamma_+(\bm{R})}e^{i\phi_-(t)-i\phi_+(t)}|_0^T\nonumber\\
		&&+\int_0^TC_2(t)[E_+(\bm{R})-E_-(\bm{R})]\frac{\braket{\psi_+^R(\bm{R})|\psi_-^R(\bm{R})}}
		{\braket{\psi_+^R(\bm{R})|\psi_+^R(\bm{R})}}
		e^{i\gamma_-(\bm{R}) -i\gamma_+(\bm{R})}e^{i\phi_-(t) -i\phi_+(t)}dt\nonumber\\
		&&-\int_0^TiC_2(t)e^{i\phi_-(t) -i\phi_+(t)}e^{i\gamma_-(\bm{R}) -i\gamma_+(\bm{R})}
		[\frac{d(\frac{\braket{\psi_+^R(\bm{R})|\psi_-^R(\bm{R})}}
			{\braket{\psi_+^R(\bm{R})|\psi_+^R(\bm{R})}}
			)}{d\bm{R}}+i\frac{d\gamma_-(\bm{R})}{d\bm{R}}-i\frac{d\gamma_+(\bm{R})}{d\bm{R}}]\frac{d\bm{R}}{dt}dt.
	\end{eqnarray}
Since $C_2(t)e^{i\gamma_-(\bm{R})-i\gamma_+(\bm{R})}e^{i\phi_-(t)-i\phi_+(t)}$ and $\frac{d\bm{R}}{dt}$ are small, $iC_2(t)\frac{\braket{\psi_+^R(\bm{R})|\psi_-^R(\bm{R})}}
{\braket{\psi_+^R(\bm{R})|\psi_+^R(\bm{R})}}
e^{i\gamma_-(\bm{R})-i\gamma_+(\bm{R})}e^{i\phi_-(t)-i\phi_+(t)}|_0^T$ and $-\int_0^TiC_2(t)e^{i\phi_-(t) -i\phi_+(t)}e^{i\gamma_-(\bm{R}) -i\gamma_+(\bm{R})}
[\frac{d(\frac{\braket{\psi_+(\bm{R})|\psi_-(\bm{R})}}
{\braket{\psi_+(\bm{R})|\psi_+(\bm{R})}}
)}{d\bm{R}}+i\frac{d\gamma_-(\bm{R})}{d\bm{R}}-i\frac{d\gamma_+(\bm{R})}{d\bm{R}}]\frac{d\bm{R}}{dt}dt$, compared with the other two terms in Eq. (\ref{intdeltaA1}), are small quantities and can be neglected. So:
\begin{eqnarray}\label{intdeltaA2}
\int_C(\tilde{\bm{A}}^{RR}_+-\bm{A}^{RR}_+)d\bm{R}&\approx&
\int_0^TC_2(t)[E_+(\bm{R})-E_-(\bm{R})]\frac{\braket{\psi_+^R(\bm{R})|\psi_-^R(\bm{R})}}
{\braket{\psi_+^R(\bm{R})|\psi_+^R(\bm{R})}}
e^{i\gamma_-(\bm{R}) -i\gamma_+(\bm{R})}e^{i\phi_-(t) -i\phi_+(t)}dt\nonumber\\
&\approx& \int_0^T[E_+(\bm{R})-E(t)]dt.
\end{eqnarray}
Consequently, Eq. (\ref{deltad0}) is satisfied when the system maintains adiabaticity.

\section*{ Appendix D: Experimental scheme in two-component Bose-Einstein condensates}

To facilitate observations in real physical systems, we will design experiments based on the bright soliton evolution in two-component Bose-Einstein condensates. We focus on the quasi-one-dimensional case, where the harmonic frequencies $\omega_y$ and $\omega_z$ ($\omega_y=\omega_z=\omega_\perp$) along the $y$ and $z$ directions are significantly larger than $\omega_x$ along the $x$ direction.
We consider the following system:
\begin{equation}\label{nonlineareq0}
	\begin{cases}
		i\hbar\frac{\partial \psi_{01}}{\partial t_0}=[-\frac{\hbar^2}{2m_1}\frac{\partial^2\psi_{01}}{\partial x_0^2}+\frac{m_1\omega_\perp}{2\pi\hbar}(g_{11}|\psi_{01}|^2+g_{12}|\psi_{02}|^2)
		+(\Delta_0+i\delta_0)]\psi_{01}+\Omega_0\psi_{02}\\
		i\hbar\frac{\partial \psi_{02}}{\partial t_0}=[-\frac{\hbar^2}{2m_2}\frac{\partial^2\psi_{02}}{\partial x_0^2}+\frac{m_2\omega_\perp}{2\pi\hbar}(g_{21}|\psi_{01}|^2+g_{22}|\psi_{02}|^2)
		+(-\Delta_0-i\delta_0)]\psi_{02}+\Omega_0^*\psi_{01}
	\end{cases},
\end{equation}
where $\delta_0$ is the gain coefficient, $-\delta_0$ is the loss coefficient, $\Delta_0$ is the detuning, and $\Omega_0$ controls the exchange of particle numbers among components. We assume that $m=m_1=m_2$ and $g_0=g_{11}=g_{22}=g_{12}=g_{21}$. After dimensionlessization, the equations are:
\begin{equation}\label{nonlineareq}
	\begin{cases}
		i\partial_t\psi_1=[-\frac{\partial_x^2}{2}+g(|\psi_1|^2+|\psi_2|^2)
		+(\Delta+i\delta)]\psi_1+\Omega\psi_2\\
		i\partial_t\psi_2=[-\frac{\partial_x^2}{2}+g(|\psi_1|^2+|\psi_2|^2)
		+(-\Delta-i\delta)]\psi_2+\Omega^*\psi_1
	\end{cases},
\end{equation}
where $t=\omega_\perp t_0$, $x=\sqrt{\frac{m\omega_\perp}{\hbar}}x_0$, $\psi_{1,2}=(\frac{\hbar}{m\omega_\perp})^{\frac{1}{4}}\psi_{01,02}$, $g=\frac{m}{2\pi\hbar^2}\sqrt{\frac{m\omega_\perp}{\hbar}}g_0$, $\Delta=\frac{\Delta_0}{\hbar\omega_\perp}$, $\delta=\frac{\delta_0}{\hbar\omega_\perp}$ and $\Omega=\frac{\Omega_0}{\hbar\omega_\perp}$.
We focus on the case where the bright soliton is wide and the nonlinear term is extremely weak. In this case, the contribution of $\partial_x^2$ can be ignored. In addition, due to the very weak nonlinearity, the approximate solutions for the system can be expressed as:
\begin{equation}\label{nonlinearsolution}
	\ket{\psi}=A\frac{\ket{\psi_{l\pm}^R}}
	{\sqrt{\braket{\psi_{l\pm}^R|\psi_{l\pm}^R}}}
	\mathrm{sech}(\frac{x}{w}),
\end{equation}
where $\psi_{l\pm}$ is the eigenstate of the linear part $H_l=\begin{pmatrix}
	\Delta+i\delta&\Omega\\
	\Omega^*&-\Delta-i\delta
\end{pmatrix}$ of the Hamiltonian in Eq. (\ref{nonlineareq}), $A$ is a constant and $w=\frac{1}{|A|\sqrt{|g|}}$ is the width of the soliton. Let $\Omega=\Omega_1-i\Omega_2$, both $\Omega_1$ and $\Omega_2$ are real numbers, by drawing an analogy between $H_l$ and Eq. (\ref{Hamiltonian}), we select the parameter space as: $(\Omega_1, \Omega_2, \Delta)$, and $\delta$ corresponds to the non-Hermitian parameter $z_0$ in Eq. (\ref{Hamiltonian}). Clearly, the eigenvalue and the eigenstate of $H_l$ are $E_{l\pm}=\pm(a_l+ib_l)$ and $
\ket{\psi^R_{l\pm}}=
\begin{pmatrix}
	\Omega_1-i\Omega_2\\
	-\Delta-i\delta+E_\pm
\end{pmatrix}$, where $a_l=\sqrt{\frac{\Omega_1^2+\Omega_2^2+\Delta^2-\delta^2
		+\sqrt{(\Omega_1^2+\Omega_2^2+\Delta^2-\delta^2)^2
			+4(\delta \Delta)^2}}{2}}$, $b_l=\frac{\delta \Delta}{a_1}$.
Given that $g$ is very small, we require the atomic scattering length to be as small as possible.
Taking $^7$Li \cite{KEStrecker,SEPollack} as an example, assuming $a_{11}\approx a_{22}\approx a_{12}\approx -0.1a_0$ ($a_0$ is Bohr radius), $A=50$, $\omega_\perp=2\pi\times 170$Hz and $\omega_x=2\pi\times 1$Hz, we estimate $g=-3.6\times 10^{-6}$, with the total particle number being $5.27\times 10^{4}$ and the full width at half maximum (FWHM) of the soliton density being $51.22\mathrm{\mu m}$.

\begin{figure}[H]
	\centering
	\includegraphics[width=11.9cm,height=4.7cm]{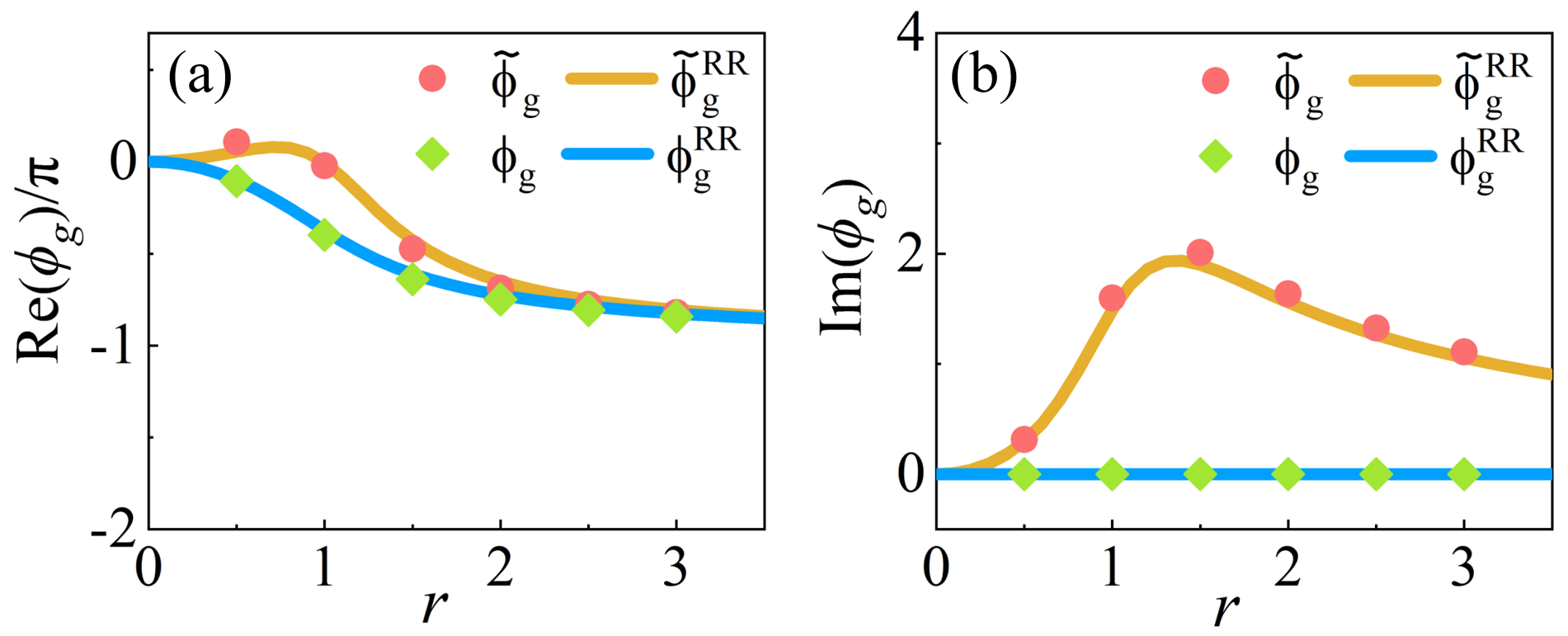}
	\caption{\label{6}(a)Real-part comparison of the geometric phase acquired by bright solitons during cyclic parameter evolution in the Eq. (\ref{nonlineareq}) system versus theoretical predictions. (b) Imaginary-part comparison of the geometric phase acquired by bright solitons during cyclic parameter evolution in the Eq. (\ref{nonlineareq}) system versus theoretical predictions. The specific information of the evolution orbit is: $\Delta=0.5$, $\Omega_1=r\cos(\omega t)$, $\Omega_2=r\sin(\omega t)$, $r=\sqrt{\Omega_1^2+\Omega_2^2}$, $\omega=0.03\pi$ and the orbit is $\omega t: 0\to 2\pi$. The non-Hermitian parameter is $Z_0=1$.  The difference of orbitals is reflected in the difference of $r$. This figure statistics the change of Berry phase when $r$ goes from 0 to 3.5.}
\end{figure}

Under the premise of satisfying the above approximations, the theoretical geometric phase $\phi_g^{RR}$ and $\tilde{\phi}_g^{RR}$ describing the adiabatic evolution of solitons can be given by $\bm{A}^{RR}_\pm$and $\bm{\tilde{A}}^{RR}_\pm$ obtained from $\ket{\psi^R_{l\pm}}$.
We take $\ket{\psi^R_{l+}}$ as an example and numerically evolve the initial state satisfying Eq. (\ref{nonlinearsolution}) under Eq. (\ref{nonlineareq}). For consistent comparison, we select the same trajectory in parameter space as in article, specifically: $\Delta=0.5$, $\Omega_1=r\cos(\omega t)$, $\delta=1$, $\Omega_2=r\sin(\omega t)$, $r=\sqrt{\Omega_1^2+\Omega_2^2}$, $\omega=0.03\pi$ and the orbit is $\omega t: 0\to 2\pi$. Note that during evolution, the imaginary part of the dynamical phase will cause amplitude growth of the soliton, thereby violating the weak nonlinearity condition. To address this, we employ a compensatory phase factor $e^{iE(t)}$ or $e^{iE_{l+}(\bm{R})}$ at each time step to eliminate the dynamical phase contribution, where $E(t)=\frac{\bra{\psi(t)}H\ket{\psi(t)}}{\braket{\psi(t)|\psi(t)}}$. When the nonlinearity is extremely weak, this procedure is valid. Under this protocol, the final phase of the evolved soliton corresponds exclusively to the geometric phases $\phi_g$ and $\tilde{\phi}_g$. We compare the theoretical predictions with the numerical results, as shown in Fig. \ref{6}. This is essentially consistent with our results obtained in linear systems.

\end{widetext}

\end{document}